\documentclass[11pt,a4paper]{article}

\usepackage{mathtools}
\usepackage{authblk} 
\usepackage{algpseudocode,algorithmicx,algorithm}

\usepackage{mathrsfs}
\usepackage{latexsym,bm}
\usepackage{amsmath,amsfonts,amssymb,amsthm,rotating}
\usepackage{extarrows}

\usepackage{graphicx,subfigure,epstopdf,float}
\usepackage{enumerate,cases,multirow}
\usepackage{makecell}
\usepackage{caption}

\usepackage{longtable,colortbl,arydshln,threeparttable}
\definecolor{mygray}{gray}{.9}

\usepackage{indentfirst}
\usepackage[top=25mm,bottom=20mm,left=25mm,right=20mm]{geometry}
\usepackage{cite}

\usepackage{listings}

\usepackage{makeidx}        
\usepackage{booktabs}
\usepackage[bookmarksnumbered,colorlinks=true,citecolor=red,linkcolor=red,hyperindex,linktocpage=true]{hyperref}

\newcommand{\ket}[1]{| #1 \rangle} 
\newcommand{\bra}[1]{\langle #1 |} 
\newcommand{\braket}[2]{ \langle  #1 | #2  \rangle }
\newcommand{\ketbra}[2]{ | #1 \rangle \langle #2 |}
\newcommand{\bb}{\boldsymbol}

\def \d {\mathrm{d}}
\def \e {\mathrm{e}}
\def \i {\mathrm{i}}

\newcounter{parentalgorithm}

\makeatother

\newtheorem{theorem}{Theorem}[section]
\newtheorem{lemma}{Lemma}[section]

\newtheorem{definition}{Definition}[section]

\theoremstyle{remark}
\newtheorem{remark}{\bf Remark}[section]

\numberwithin{equation}{section}

\begin{document}

\title{Quantum Circuits for the heat equation with physical boundary conditions via
Schr\"odingerisation}

\author[1]{Shi Jin\thanks{shijin-m@sjtu.edu.cn}}
\author[1, 2]{Nana Liu\thanks{nana.liu@quantumlah.org}}
\author[3]{Yue Yu\thanks{terenceyuyue@xtu.edu.cn}}
\affil[1]{School of Mathematical Sciences, Institute of Natural Sciences, MOE-LSC, Shanghai Jiao Tong University, Shanghai, 200240, China}
\affil[2]{University of Michigan-Shanghai Jiao Tong University Joint Institute, Shanghai 200240, China}
\affil[3]{School of Mathematics and Computational Science, Hunan Key Laboratory for Computation and Simulation in Science and Engineering, Key Laboratory of Intelligent Computing and Information Processing of Ministry of Education, National Center for Applied Mathematics in Hunan, Xiangtan University, Xiangtan, Hunan 411105, PR China}

\maketitle

\begin{abstract}
  This paper explores the explicit design of quantum circuits for quantum simulation of partial differential equations (PDEs) with physical boundary conditions. These equations and/or their discretized forms usually do not evolve via unitary dynamics,
  thus are not suitable for quantum simulation.  Boundary conditions (either time-dependent or independent) make the problem more difficult.
  To tackle this challenge, the Schr\"odingerisation method can be employed, which converts linear partial and ordinary differential equations with non-unitary dynamics into systems of Schr\"odinger-type equations, via the so-called warped phase transformation that maps the equation into one higher dimension.
  Despite advancements in Schr\"odingerisation techniques, the explicit implementation of quantum circuits for solving general PDEs, especially with physical boundary conditions, remains underdeveloped.
  We present two methods for handling the inhomogeneous terms arising from time-dependent physical boundary conditions.
  One approach utilizes Duhamel's principle to express the solution in integral form and employs linear combination of unitaries (LCU) for coherent state preparation.
  Another method applies an augmentation  to transform the inhomogeneous problem into a homogeneous one. We then apply the quantum simulation technique from \cite{CJL23TimeSchr} to transform the resulting non-autonomous system to an autonomous system in one higher dimension. We provide detailed implementations of these two methods and conduct a comprehensive complexity analysis in terms of queries to the time evolution input oracle.
\end{abstract}


\tableofcontents

\section{Introduction}

Partial differential equation (PDE) serve as essential tools for investigating the dynamic behaviors of numerous physical systems, with applications spanning combustion, atmospheric and ocean circulation, and electromagnetic waves. Simulating such complex dynamical systems using reasonable computational time and resources is crucial. Despite substantial progress facilitated by high-performance computing clusters, managing the large computational overhead remains a challenge, especially in high-dimensional spaces (such as the $N$-body Schr\"odinger equation and kinetic equations like the Boltzmann equation), or when dealing with multiple time and space scales in high frequency waves and turbulence.

One of the primary challenges in classical numerical simulations of PDEs is the curse of dimensionality, particularly evident in high-dimensional PDEs in statistical physics and machine learning applications. Recently, there has been growing interest in developing quantum algorithms~--~that use quantum computers that might be available in the future~--~to solve PDEs \cite{Cao2013Poisson,Berry-2014,qFEM-2016,Costa2019Wave,Engel2019qVlasov,Childs-Liu-2020,
Linden2020heat,Childs2021high,JinLiu2022nonlinear,GJL2022QuantumUQ,JLY2022multiscale}. Many of these algorithms leverage the exponential speedup advantages in solving quantum linear systems of equations \cite{HHL2009,Childs2017QLSA,Costa2021QLSA,Berry-2014,BerryChilds2017ODE,Childs-Liu-2020,Subasi2019AQC}.
A strategy for crafting quantum PDE solvers involves discretizing spatial variables to formulate a system of ordinary differential equations (ODEs), which can then be addressed using quantum ODE solvers \cite{Berry-2014,BerryChilds2017ODE,Childs-Liu-2020}. Importantly, if the solution operator for the resulting ODE system is unitary, quantum simulations can be executed with reduced time complexity compared to other quantum linear algebra methods, such as quantum difference methods \cite{Berry-2014,JLY2022multiscale}. In cases where the system is not unitary, it is necessary to ``dilate'' it to a unitary system \cite{JLY22SchrLong,Javier2022optionprice,Burgarth23dilations,CJL23TimeSchr}.

Among the unitarization approaches, the {\it Schr\"odingerisation} method introduced in \cite{JLY22SchrShort, JLY22SchrLong, analogPDE}, is a  simple and generic framework that allows quantum simulation for {\it all} linear PDEs and ODEs. The idea is to use a warped phase transform that maps the equations to one higher dimension, which, in the Fourier space, become a system of Schr\"odinger type equations!
The approach has been expanded to address a wide array of problems, including open quantum systems within bounded domains with non-unitary artificial boundary conditions \cite{JLLY23ABC}, problems entailing physical boundary or interface conditions \cite{JLLY2024boundary}, Maxwell's equations \cite{JLC23Maxwell}, the Fokker-Planck equation \cite{JLY24FokkerPlanck}, ill-posed scenarios such as the backward heat equation \cite{JLM24SchrBackward}, linear dynamical systems with inhomogeneous terms \cite{JLM24SchrInhom}, non-autonomous ordinary and partial differential equations \cite{CJL23TimeSchr}, iterative linear algebra solvers \cite{JinLiu-LA}, etc. This approach is also natural for continuous variables \cite{analogPDE} thus is also an idea~--~so far the only possible one~--~for the analog quantum simulation of PDEs and ODEs. This method can also be extended to parabolic partial differential equations using a Jaynes-Cummings-like model -- which is more widely experimentally available -- as was proposed in \cite{JL24JaynesCummings}.

In this paper, we focus on designing explicit quantum circuits for simulating PDEs with physical boundary conditions. Recent work by Sato et al. \cite{Sato24Circuit} introduced a novel approach to construct scalable quantum circuits tailored for wave or Schr\"odinger-type PDEs, utilizing the Bell basis to diagonalize each Hamiltonian term.  This methodology has been further extended  in \cite{HuJin24SchrCircuit} to develop quantum circuits for general PDEs, which may not strictly adhere to unitary dynamics, through the application of the Schr\"odingerisation technique. However, these studies do not comprehensively address problems involving physical boundary conditions.
In practical applications, solving PDEs within bounded domains necessitates specifying boundary conditions for the problem to be well-defined. Common physical boundary conditions include Dirichlet, Neumann, periodic, and Robin (mixed) conditions, which do not exhibit unitary properties and thus pose challenges for quantum simulations.  Although these issues have been explored in \cite{JLLY2024boundary,JLLY23ABC} using the Schr\"odingerisation technique, our paper aims to explicitly tackle the design of quantum circuits for solving such problems with the quantum circuit in \cite{HuJin24SchrCircuit} as an input oracle for time evolution. We provide a concise review of the quantum circuit construction for the Hamiltonian evolution operator proposed in \cite{HuJin24SchrCircuit}, address treatments for physical boundary conditions and discuss modifications to both the gate construction and the time complexity.
In particular, we achieve exponential speedup with respect to the discretization of the auxiliary variable in the Schr\"odingerisation approach through careful consideration of the select oracle.
Specifically, the approximate evolution operator $V_{\text{heat}}(\tau)$ in \cite{HuJin24SchrCircuit} is a select oracle which can be realized by performing the controlled operator $c-\tilde{V}_0^{2^m}(\tau)$, where the complexity of the quantum circuit is evaluated based on the count of controlled $\tilde{V}_0(\tau)$ gates, resulting in $\sum_{m=0}^{n_p-1} 2^m = \mathcal{O}(2^{n_p})$ uses of $\tilde{V}_0(\tau)$. We present a modified but equivalent circuit in Fig.~\ref{fig:Vheatm1} for $V_{\text{heat}}(\tau)$, with $c-\tilde{V}_0^{2^m}(\tau)$ replaced by $c-\tilde{V}_0(2^m\tau)$ since $ \tilde{V}_0^{2^m}(\tau) = \tilde{V}_0(2^m\tau)$. The significant reduction in complexity within one segment arises from the fact that $\tilde{V}_0(2^m\tau)$ uses the same number of gates regardless of the parameter of $\tilde{V}_0(s)$.
While we specifically discuss the heat equation here, the approach can be generalized to other scenarios, such as the advection equation with inflow boundary conditions, the Fokker-Planck equation with no-flux boundaries \cite{Hu2023PreservingFP}, quantum dynamics incorporating artificial boundary conditions \cite{JLLY23ABC}, and problems involving interface conditions \cite{JLLY2024boundary}.

The paper is structured as follows:
In Section \ref{sec:Schrodingerization}, we provide an overview of the Schr\"odingerisation approach and present two methods for addressing the inhomogeneous terms resulting from time-dependent physical boundary conditions. One approach utilizes Duhamel's principle to express the solution as an integral form and employs linear combination of unitaries (LCU) to coherently prepare the state. An alternative method applies the augmentation technique described in \cite{JLY22SchrLong, JLLY2024boundary, JLM24SchrInhom} to transform the inhomogeneous problem into a homogeneous one. We then apply the quantum simulation technique from \cite{CJL23TimeSchr} to handle the resulting non-autonomous system, transferring it to an autonomous one in one higher dimension. This approach avoids the use of the computationally complicated Dyson's series.
Section \ref{sec:representation} discusses the treatment of boundary conditions using shift operators and provides details on the decomposition of these operators as outlined in \cite{Sato24Circuit}. Unlike the original finite difference operator representation in \cite{Sato24Circuit}, here we focus solely on representations in terms of shift operators, which align well with finite difference discretizations.
In Section \ref{sec:SchrLap}, we briefly review the construction of the quantum circuit for the Hamiltonian evolution operator proposed in \cite{HuJin24SchrCircuit}, addressing treatments for physical boundary conditions.
Sections \ref{sec:LCU} and \ref{sec:autonomization} focus on the two approaches for handling the inhomogeneous term. We provide detailed implementation of these methods and conduct a thorough complexity analysis in terms of queries to the time evolution input oracle.
Conclusions are presented in the final section.

\section{Schr\"odingerisation method for linear differential equations} \label{sec:Schrodingerization}

For a linear partial differential equation with constant coefficients, numerical methods can be employed for spatial variables to obtain a linear system of differential equations in the form:
\begin{equation}\label{ODElinear}
\begin{cases}
 \dfrac{\d \bb{u}(t)}{\d t} = A \bb{u}(t) + \bb{f}(t),\\
 \bb{u}(0) = \bb{u}_0,
\end{cases}
\end{equation}
where $\bb{u} = [u_0, u_1,\cdots, u_{N-1}]^T$ and $A \in \mathbb{C}^{N\times N}$. The time-dependence of the inhomogeneous term $\bb{f}(t) = [f_0(t),f_1(t),\cdots,f_{N-1}(t)]^T$ arises from the application of the boundary conditions.

For a general $A $,
we first decompose $A$ into a Hermitian term and an anti-Hermitian term:
\[A = H_1 + \i H_2, \qquad \i = \sqrt{-1},\]
where
\[
H_1 = \frac{A+A^{\dagger}}{2} = H_1^{\dagger}, \qquad H_2 = \frac{A-A^{\dagger}}{2 \i} = H_2^{\dagger}.
\]
Here ``$\dagger$'' denotes conjugate transpose.
A natural assumption is that the semi-discrete system \eqref{ODElinear} inherits the stability of the original PDE, which implies that $H_1$ is negative semi-definite.
Using the warped phase transformation $\bb{v}(t,p) = \e^{-p} \bb{u}(t)$ for $p\ge 0$ and symmetrically extending the initial data to $p<0$,  system \eqref{ODElinear} is  then transformed to a system of linear convection equations \cite{JLY22SchrShort,JLY22SchrLong}:
\begin{equation}\label{u2v}
\begin{cases}
 \frac{\d}{\d t} \bb{v}(t,p) = A \bb{v}(t,p) = - H_1 \partial_p \bb{v} + \i H_2 \bb{v} + \bb{g}(t,p), \\
 \bb{v}(0,p) = \e^{-|p|} \bb{u}_0,
 \end{cases}
\end{equation}
where $ \bb{g}(t,p) = \e^{-p} \bb{f}(t)$. When $H_1$ is negative semi-definite, $\bb{u}(t)=\e^p \bb{v}(t,p)$ for all $p>0$, so we can restore the solution $\bb{u}(t)$ by simply choosing any $p = p_{k_*}>0$.

Let $p\in [L,R]$ with $L<0$ and $R>0$ satisfying $\e^{-L} \approx 0$ and $\e^R \approx 0$. Then one can apply the periodic boundary condition in the $p$ direction and use the Fourier spectral method by discretizing the $p$ domain.  Toward this end,  we choose a uniform mesh size $\Delta p = (R-L)/N_p$ for the auxiliary variable with $N_p=2^{n_p}$ being an even number, with the grid points denoted by $a = p_0<p_1<\cdots<p_{N_p} = b$.
To compute $\bb v(t,p),$ let the vector $\bb{w}$ be the collection of the function $\bb v$ at these grid points, defined more precisely as follows,
\begin{equation}\label{orderw}
\bb{w} = [\bb{w}_1; \bb{w}_2; \cdots; \bb{w}_n],
\end{equation}
with ``;'' indicating the straightening of $\{\bb{w}_i\}_{i\ge 1}$ into a column vector. This can also be expressed as a superposition state using $\ket{k}$ as a new basis,
\[
 \bb{w}_i = \sum_{k=0}^{N_p-1} \bb{v}_i (t,p_k) \ket{k},
\]
where $\bb{v}_i$ is the $i$-th entry of $\bb{v}$.

By applying the discrete Fourier transformation in the $p$  direction, one arrives at
\begin{equation}\label{heatww}
\frac{\d}{\d t} \bb{w}(t) = -\i ( H_1 \otimes P_\mu ) \bb{w} + \i (H_2 \otimes I) \bb{w} + \bb{b}(t), \quad
\bb{w}(0) = \bb{u}_0 \otimes [\e^{-|p_0|}, \cdots, \e^{-|p_{N_p-1}|}]^T,
\end{equation}
with $\bb{b}(t) = \sum\limits_{i,k} \bb{g}_i(t,p_k) \ket{i,k}$ and $\bb{g}_i$ being the $i$-th entry of $\bb{g}$.
 Here, $P_\mu$ is the matrix expression of the momentum operator $-\i\partial_p$, given by
\begin{equation}\label{Pmu}
P_\mu = \Phi D_\mu \Phi^{-1},  \qquad D_\mu = \text{diag}(\mu_{-N}, \cdots, \mu_{N-1}),
\end{equation}
where $\mu_l = 2\pi l/(R-L)$ are the Fourier modes and
\[\Phi = (\phi_{jl})_{M\times M} = (\phi_l(x_j))_{N_p\times N_p}, \qquad \phi_l(x) = \e ^{\i \mu_l (x-L)} \]
is the matrix representation of the discrete Fourier transform.
At this point, we have successfully mapped the dynamics back to a Hamiltonian system.
By a change of variables $\tilde{\bb{w}} = (I \otimes \Phi^{-1})\bb{w}$, one has
\begin{equation}\label{generalSchr}
\frac{\d}{\d t} \tilde{\bb{w}}(t) = -\i ( H_1 \otimes D_\mu ) \tilde{\bb{w}} + \i (H_2 \otimes I) \tilde{\bb{w}} + \tilde{\bb{b}}(t)
= \i H \tilde{\bb{w}}(t) + \tilde{\bb{b}}(t),
\end{equation}
where
\[ H = - H_1 \otimes D_\mu + H_2 \otimes I, \qquad  \tilde{\bb{b}} = (I \otimes \Phi^{-1})\bb{b}.
\]

With the inhomogeneous term $\tilde{\bb{b}}$, there are two approaches. First, one can use Duhamel's principle to express the solution as
\begin{equation}\label{Duhamel}
\tilde{\bb{w}}(t) = \e^{\i H t } \tilde{\bb{w}}(0) + \int_0^t \e^{\i H(t-s) } \tilde{\bb{b}}(s) \d s
\end{equation}
and utilize the LCU to coherently prepare the state $\tilde{\bb{w}}(t)$.

 An alternative approach is to employ the augmentation technique in \cite{JLY22SchrLong,JLLY2024boundary,JLM24SchrInhom} to transform equation \eqref{ODElinear} to the homogeneous case where $\bb{f}(t) = \bb{0}$. Denote $F_\varepsilon(t) = \text{diag}(\tilde{\bb{f}_\varepsilon}(t))$ as a diagonal matrix with the $i$-th entry of $\tilde{\bb{f}_\varepsilon}(t)$ given by
 \[\tilde{f}_{\varepsilon,i}(t) = \frac{f_i(t)}{( (f_i^2)_{ave} + \varepsilon^2 )^{1/2}}, \qquad i=0,1,\cdots,N-1,\]
 where
 \[
 (f_i^2)_{ave} := \frac{1}{T} \int_0^T |f_i(t)|^2 \d t.
 \]
 $\varepsilon = 1/\sqrt{N}$ is included to prevent division by zero in the denominator. Its size is  chosen so the error introduced here is the same order as that for the discrete Fourier transform for a function that is not continuous differentiable such as $\e^{-|p|}$.

 Then one can rewrite \eqref{generalSchr} as an enlarged system
\[\frac{\d }{\d t} \bb{u}_{\varepsilon}(t)
= \begin{bmatrix}
A  &  F_\varepsilon(t) \\
O     &  O
\end{bmatrix}\bb{u}_{\varepsilon}(t), \qquad \bb{u}_{\varepsilon}(t) = \begin{bmatrix}
\bb{u}(t) \\
\bb{r}_\varepsilon(t)
\end{bmatrix}, \qquad
\bb{u}_{\varepsilon}(0) = \begin{bmatrix}
\bb{u}_0 \\
\bb{r}_\varepsilon(0)
\end{bmatrix},
\]
where $\bb{r}_\varepsilon(t)$ is a constant column vector with the $i$-th entry given by
\[r_{\varepsilon,i}(t) = ( (f_i^2)_{ave} + \varepsilon^2 )^{1/2}, \qquad i=0,1,\cdots, N-1.\]
For this new system, the probability of projecting onto $\ket{\bb{u}(t)}$ is
\[\text{Pr} = \frac{\|\bb{u}(t)\|^2}{\|\bb{u}_\varepsilon(t)\|^2}
= \frac{\|\bb{u}(t)\|^2}{\|\bb{u}(t)\|^2 + \|\bb{f}\|_{ave}^2+1},\]
with $\|\bb{f}\|_{ave} = \Big(\sum\limits_{i=0}^{N-1} (f_i^2)_{ave}\Big)^{1/2}$.

Applying the Schr\"odingerisation technique reviewed above, one gets
\begin{equation}\label{generalSchrEnlarge}
\frac{\d}{\d t} \tilde{\bb{w}}_{\varepsilon}(t,p) = \i H_{\varepsilon}(t) \tilde{\bb{w}}_{\varepsilon}(t,p), \qquad
\tilde{\bb{w}}_{\varepsilon}(0,p) = \tilde{\bb{w}}_{\varepsilon,0} = \begin{bmatrix}
\bb{u}_0 \\
\bb{r}_\varepsilon(0)
\end{bmatrix} \otimes \Phi^{-1} [\e^{-|p_0|}, \cdots, \e^{-|p_{N_p-1}|}]^T,
\end{equation}
where $\tilde{\bb{w}}_{\varepsilon} = (I\otimes \Phi^{-1}) \bb{w}_{\varepsilon}$ with $\bb{w}_{\varepsilon}$ defined as $\bb{w}$ in \eqref{orderw},
\[H_{\varepsilon}(t) = - H_{1,\varepsilon}(t) \otimes D_\mu + H_{2,\varepsilon}(t) \otimes I, \]
\[H_{1,\varepsilon}(t) = \begin{bmatrix}
H_1  &  F_\varepsilon (t)/2 \\
F_\varepsilon (t)/2     &  O
\end{bmatrix}, \qquad H_{2,\varepsilon}(t) = \begin{bmatrix}
H_2  &  F_\varepsilon(t)/2\i \\
-F_\varepsilon(t)/2\i     &  O
\end{bmatrix}.\]
Since the evolution matrix is time-dependent, we can apply the quantum simulation technique in \cite{CJL23TimeSchr} for non-autonomous systems, where a non-autonomous system is transferred to an autonomous one in one higher dimension. This approach avoids the use of the complicated Dyson's series. Details will be given in the subsequent section.

\begin{remark} \label{rem:plus}
Note that the maximum eigenvalue of $H_{1,\varepsilon}$ satisfies
\[\lambda_{\max}(H_{\varepsilon}(t)) \le \lambda_{\max}(H_1) + \frac12 \max_{i} \frac{|f_i(t)|}{( (f_i^2)_{ave} + \varepsilon^2 )^{1/2}} = : \lambda_\varepsilon(t).\]
Some of the eigenvalues may become positive so one needs to be careful when recovering $\bb{w}_\varepsilon(t)$.
According to Theorem 3.1 of \cite{JLM24SchrInhom}, the solution can be recovered by $\bb{w}_\varepsilon(t) = \e^p \tilde{\bb{w}}_\varepsilon(t,p)$ for $p\ge \lambda_+ t$, where $\lambda_+ = \max\{ \lambda_\varepsilon(t), 0\}$. If $f_i(t)$ is not significantly different from the square root of $(f_i^2)_{ave}$, one can expect $\lambda_+ = \mathcal{O}(1)$.
\end{remark}

\section{Representation of discrete Laplacian with physical boundary conditions} \label{sec:representation}

In a recent study by Sato et al. \cite{Sato24Circuit}, a novel approach was introduced for the construction of scalable quantum circuits tailored specifically for wave or Schr\"odinger-type partial differential equations (PDEs), where the Bell basis is employed to diagonalize each term of the Hamiltonian. This methodology has been further extended  in \cite{HuJin24SchrCircuit} to develop quantum circuits for heat and advection equations, which may not strictly adhere to unitary dynamics, through the application of the Schr\"odingerisation technique.
In these studies, the treatment of boundary conditions is not discussed in detail. This section aims to elaborate on this issue for the following heat equation
\[\partial_ t u(t,x) = \Delta u(t,x), \qquad x \in \mathbb{R}^d, \qquad u(0,x) = u_{in}(x).\]

\subsection{Finite difference discretization with physical boundary conditions}\label{subsec:FD}

\subsubsection{The Dirichlet boundary condition}

We begin by considering the one-dimensional case in order to establish the definition of the shift operators.  We take the domain $\Omega = (a,b)$ and impose the inhomogeneous Dirichlet boundary conditions
\[u(t,x=a) = u_a(t), \qquad  u(t,x=b) = u_b(t).\]

The domain is uniformly divided into $M$ intervals of length $\Delta x = (b-a)/M$. Let $u_i(t) = u(t,x_i)$ be the evaluation of $u$ at $x_i = a + i \Delta x$ with $i=0,1,\cdots,M$. The unknown values are $u_1(t), \cdots, u_{M-1}(t)$. For the spatial discretization, we apply the central difference discretization at $x = x_j$ for $j = 1,\cdots,M-1$:
\begin{equation}\label{heatdiscretisation}
\frac{{\rm d}}{{\rm d}t}u_j(t) = \frac{u_{j-1}(t) - 2u_j(t) + u_{j+1}(t)}{\Delta x^2},  \quad j = 1,\cdots, M-1.
\end{equation}
Let $\bb{u}(t) = [u_1(t), \cdots, u_{M-1}(t)]^T = \sum_{j=0}^{M-2} u_{j+1} \ket{j}$, where $\ket{j} = [0, 0, \cdots, 0, 1, 0, \cdots, 0]^T \in \mathbb{R}^{M-1}$ is a column vector with the $j$-th entry being 1. One gets the system \eqref{ODElinear} with
\[A = \frac{1}{\Delta x^2}
\begin{bmatrix}
-2  &  1       &           &      &    \\
 1  & -2       & \ddots    &      &    \\
    &  \ddots  & \ddots    &  \ddots    &    \\
    &          & \ddots    & -2   & 1  \\
    &          &           &  1   & -2 \\
\end{bmatrix}_{(M-1) \times (M-1)}
, \qquad
\bb{f}(t) = \frac{1}{\Delta x^2}
\begin{bmatrix}
u_0(t) \\
0  \\
\vdots\\
0 \\
u_M(t) \\
\end{bmatrix}.
\]
This matrix form can also be written as
\[
\frac{\d }{\d t} \bb{u}(t) = \frac{1}{\Delta x^2}( \bb{u}^- - 2 \bb{u} + \bb{u}^+),
\]
where
\begin{align*}
& \bb{u}^- = \begin{bmatrix}
u_0 \\ u_1 \\ \vdots \\ u_{M-3} \\ u_{M-2}
\end{bmatrix} =
\begin{bmatrix}
0    &         &        &    &    \\
1    & 0       &        &    &   \\
     & \ddots  & \ddots &    &   \\
     &         &    1    &  0  &   \\
     &         &        & 1  & 0
\end{bmatrix}\begin{bmatrix}
u_1 \\ u_2 \\ \vdots \\ u_{M-2} \\ u_{M-1}
\end{bmatrix} + \begin{bmatrix}
u_0 \\ 0 \\ \vdots \\ 0 \\ 0
\end{bmatrix}  =: S^+ \bb{u} + u_0 \ket{0}, \\
& \bb{u}^+ = \begin{bmatrix}
u_2 \\ u_3 \\ \vdots \\ u_{M-1} \\ u_M
\end{bmatrix} =
\begin{bmatrix}
0    & 1   &         &          &    \\
     & 0   &  1      &          &   \\
     &     & \ddots  &  \ddots  &   \\
     &     &         &  0       & 1  \\
     &     &         &          & 0
\end{bmatrix}\begin{bmatrix}
u_1 \\ u_2 \\ \vdots \\ u_{M-2} \\ u_{M-1}
\end{bmatrix} + \begin{bmatrix}
0 \\ 0 \\ \vdots \\ 0 \\ u_M
\end{bmatrix}  =: S^- \bb{u} + u_M \ket{M-2},
\end{align*}
and the shift operators $S^+$ and $S^-$ satisfy
\begin{align*}
& S^+\ket{j} = \ket{j+1}, \qquad j = 0,1,\cdots,M-3, \qquad S^+ \ket{M-2}= \bb{0}, \\
& S^-\ket{j} = \ket{j-1}, \qquad  j = 1,2,\cdots, M-2, \qquad S^- \ket{0} = \bb{0},
\end{align*}
where $\ket{j}$ is the computational basis state for $j = 0,1,\cdots, M-2$.
The heat equation with the Dirichlet boundary conditions is then represented by the shift operators as
\[\frac{\d }{\d t}  \bb{u}(t) = A \bb{u}(t) + \bb{f}(t), \qquad
\bb{u}(0) = [u_{in}(x_1), \cdots, u_{in}(x_{M-1})]^T,\]
where
\[A = D_D^\Delta = \frac{S^+ - 2 I^{\otimes n_x} + S^-}{\Delta x^2},\]
where $M-1 = 2^{n_x} = N_x$.

For $d$ dimensions, the domain is taken as $\Omega = (a,b)^d$. To construct a spatial discretization, we  introduce $M+1$ spatial mesh points $a = x_{i,0}<x_{i,1}<\cdots<x_{i,M} = b$ by $x_{i,j} = a+j \Delta x$ in the $x_i$-direction, where $\Delta x = (b-a)/M$. Let $\bb{j} = (j_1,j_2,\cdots, j_d)$, where $j_k=0,1,\cdots,M$ for $k=1,\cdots,d$. With the central difference applied, we get
\[\frac{\d }{\d t} u_{\bb{j}}(t) =  \sum\limits_{k=1}^d \frac{ u_{\bb{j}-\bb{e}_k}(t)-2 u_{\bb{j}}(t) + u_{\bb{j}+\bb{e}_k}(t)}{\Delta x} ,\]
where $\bb{e}_k = (0,\cdots,0,1,0,\cdots,0)$ with the $k$-th entry being 1 and $j_k = 1,2,\cdots, M-1$.  Denote by $\bb{u}$ the vector form of the $d$-order tensor $(u_{\bb{j}}) = (u_{j_1,j_2,\cdots, j_d})$:
\[\bb{u} = \sum\limits_{\bb{j}} u_{\bb{j}} \ket{\bb{j}}
= \sum\limits_{j_1,j_2,\cdots,j_d = 0}^{M-2} u_{j_1+1,j_2+1,\cdots,j_d+1} \ket{j_1,j_2,\cdots,j_d}.\]
The associated linear system can be represented as
\[\sum\limits_{\bb{j}} \frac{\d }{\d t} u_{\bb{j}}(t) \ket{\bb{j}} =  \sum\limits_{\bb{j}} \sum\limits_{k=1}^d \frac{ u_{\bb{j}-\bb{e}_k}(t)-2 u_{\bb{j}}(t) + u_{\bb{j}+\bb{e}_k}(t)}{\Delta x^2}  \ket{\bb{j}}.\]
To write the above system in matrix form, let us assume that $u_{\bb{j}}$ can be decomposed as $u_{\bb{j}} = u_{j_1} u_{j_2} \cdots u_{j_d}$.
Noting that $u_{j_1, \cdots, j_d}$ ($j_k = 0, M$) are the given boundary values in the $x_k$ direction, we have
\begin{align*}
\sum\limits_{\bb{j}} u_{\bb{j}-\bb{e}_k}\ket{\bb{j}}
& = \sum\limits_{j_1,j_2,\cdots,j_d = 0}^{M-2} u_{j_1+1,\cdots,j_k, \cdots, j_d+1} \ket{j_1,\cdots,j_k,\cdots,j_d} \\
& = \sum\limits_{j_i = 0, i \ne k}^{M-2} u_{j_1+1,\cdots,0, \cdots, j_d+1} \ket{j_1,\cdots,0,\cdots,j_d} \\
& \quad + \sum\limits_{j_1 = 0}^{M-2} u_{j_1+1} \ket{j_1} \otimes \cdots  \otimes \sum\limits_{j_k = 1}^{M-2} u_{j_k}\ket{j_k} \otimes  \cdots \otimes \sum\limits_{j_d = 0}^{M-2} u_{j_d+1}\ket{j_d} \\
& = : \bb{u}_{0k} + \bb{u}^{(1)} \otimes \cdots \otimes S^+ \bb{u}^{(k)} \otimes \cdots \otimes \bb{u}^{(d)} \\
& = \bb{u}_{0k} + ( I^{\otimes n_x} \otimes \cdots \otimes S^+  \otimes \cdots \otimes I^{\otimes n_x}) (\bb{u}^{(1)} \otimes \cdots \otimes \bb{u}^{(d)})\\
& = \bb{u}_{0k} + ( I^{\otimes n_x} \otimes \cdots \otimes S^+  \otimes \cdots \otimes I^{\otimes n_x}) \bb{u},
\end{align*}
where
\[\bb{u}_{0k}(t) = \sum\limits_{j_i = 0, i \ne k}^{M-2} u_{j_1+1,\cdots,0, \cdots, j_d+1} \ket{j_1,\cdots,0,\cdots,j_d} \quad (\mbox{0 is located at the $k$-th position})\]
is the vector generated by left boundary values in the $x_k$-direction,
\[\bb{u}^{(i)} = \sum\limits_{j_i = 0}^{M-2} u_{j_i+1} \ket{j_i}, \qquad i = 1,2,\cdots,d.\]
Similarly, one has
\begin{align*}
\sum\limits_{\bb{j}} u_{\bb{j} + \bb{e}_k}\ket{\bb{j}}
& = \sum\limits_{j_1,j_2,\cdots,j_d = 0}^{M-2} u_{j_1+1,\cdots,j_k+2, \cdots, j_d+1} \ket{j_1,\cdots,j_k,\cdots,j_d} \\
& = \bb{u}_{Mk} + ( I^{\otimes n_x} \otimes \cdots \otimes S^-  \otimes \cdots \otimes I^{\otimes n_x}) \bb{u},
\end{align*}
with
\[\bb{u}_{Mk}(t) = \sum\limits_{j_i = 0, i \ne k}^{M-2} u_{j_1+1,\cdots,M, \cdots, j_d+1} \ket{j_1,\cdots,M-2,\cdots,j_d}.\]
We therefore obtain the system \eqref{ODElinear} with
\begin{equation}\label{heatmatrixd}
A = \bb{D}_D^\Delta =: \sum_{\alpha = 1}^d (D_D^\Delta)_\alpha,
\end{equation}
where
\[(\bullet)_\alpha := I^{\otimes (d-\alpha) n_x} \otimes \bullet \otimes I^{\otimes (\alpha-1) n_x},\]
and
\[\bb{f}(t) = \frac{1}{\Delta x^2} \sum\limits_{k=1}^d ( \bb{u}_{0,k} + \bb{u}_{M,k} ).\]

\subsubsection{The Neumann boundary condition}

We  consider the heat equation with the mixed boundary conditions:
\[u(t,a) = g(t), \qquad u_x(t,b) = h(t). \]
The discretisation at the interior node is still given by \eqref{heatdiscretisation}. For the right boundary, we introduce a ghost point $x_{M+1} = x_M + \Delta x$ and use the central difference to discretize the first-order derivative,
\[\frac{u_{M+1}(t) - u_{M-1}(t)}{2 M} = h(t).\]
To get a closed system, we assume the discretisation in \eqref{heatdiscretisation} is valid at $x = x_M$:
\[\frac{{\rm d}}{{\rm d}t}u_M(t) = \frac{u_{M-1}(t) - 2u_M(t) + u_{M+1}(t)}{\Delta x^2}.\]
Eliminating the ghost values to get
\[\frac{{\rm d}}{{\rm d}t}u_M(t) = \frac{2 u_{M-1}(t) - 2u_M(t) }{\Delta x^2} + \frac{2h(t) }{\Delta x},  \quad j = M.\]
Let $\bb{u}(t) = [u_1(t), u_2(t), \cdots, u_M(t)]^T$. Then one gets the system \eqref{ODElinear} with
\[A = D_N^\Delta = \frac{1}{\Delta x^2}
\begin{bmatrix}
-2  &  1       &           &          &                 \\
 1  & -2       & \ddots    &          &                 \\
    &  \ddots  & \ddots    &  \ddots  &               \\
    &           &  1       & -2       &   1     \\
    &           &          & 1       & -2      \\
\end{bmatrix}_{M \times M}
, \qquad
\bb{f}(t) = \frac{1}{\Delta x^2}
\begin{bmatrix}
g(t) \\
0  \\
\vdots\\
0 \\
2 h(t) \Delta x
\end{bmatrix}.
\]

The coefficient matrix $D_N^\Delta$ maintains the same form as for the Dirichlet boundary conditions in $d$ dimensions. The right-hand side vector can be derived in a similar manner. The slight difference lies in the implementation, where we should set $M = 2^{n_x} = N_x$ and take
\[D_N^\Delta = \frac{S^+ - 2 I^{\otimes n_x} + S^-}{\Delta x^2}, \]
where $S^+$ and $S^-$ are $n_x$-qubit matrices.

\subsubsection{The periodic boundary condition}

We now impose periodic boundary condition in one dimension:
\[u(t, x = a) = u(t, x = b).\]
The discretisation in \eqref{heatdiscretisation} remains valid for $j = 1,2,\cdots,M-1$. For $j = 0$,  we introduce a ghost point $x_{-1} = x_0 - \Delta x$ and assume the discretisation is applicable at $x = x_0$:
\[
\frac{{\rm d}}{{\rm d}t}u_0(t) = \frac{u_{M-1}(t) - 2u_0(t) + u_1(t)}{\Delta x^2},  \quad j = 0,
\]
where $u_{-1} = u_{M-1}$ has been employed to maintain periodicity. For $j = M-1$, the periodic boundary condition $u_0 = u_M$ gives
\[
\frac{{\rm d}}{{\rm d}t}u_0(t) = \frac{u_{M-2}(t) - 2u_{M-1}(t) + u_0(t)}{\Delta x^2},  \quad j = M-1.
\]
Let $\bb{u}(t) = [u_0(t), u_1(t), \cdots, u_{M-1}(t)]^T$. We then obtain the system \eqref{ODElinear} with
\[A = D_P^\Delta = \frac{1}{\Delta x^2}
\begin{bmatrix}
-2  &  1       &           &          &       1          \\
 1  & -2       & \ddots    &          &                 \\
    &  \ddots  & \ddots    &  \ddots  &               \\
    &           &  1       & -2       &   1     \\
 1   &           &          & 1       & -2      \\
\end{bmatrix}_{M \times M}
, \qquad
\bb{f}(t) = \bb{0},
\]
where $D_P^\Delta$ can be written as
\begin{align}
D_P^\Delta
& = \frac{S^+ - 2 I^{\otimes n_x} + S^-}{\Delta x^2} + \frac{1}{\Delta x^2} (\ket{0}\bra{M-1} + \ket{M-1}\bra{0}) \nonumber\\
& = \frac{S^+ - 2 I^{\otimes n_x} + S^-}{\Delta x^2} + \frac{1}{\Delta x^2} (\sigma_{01}^{\otimes n_x} + \sigma_{10}^{\otimes n_x}) , \label{DPDelta}
\end{align}
and $S^+$ and $S^-$ are $n_x$-qubit matrices and
\[
    \sigma_{01}  = |0\rangle \langle 1|, \qquad
    \sigma_{10} = |1\rangle \langle 0|.
\]

\subsection{Decomposition of the shift operators}

Ref.~\cite{Sato24Circuit} presented an essential decomposition of the shift operators. Here we provide a detailed explanation of this decomposition.

Let $N = 2^n$, where $n$ is the number of qubits. Then the $n$-qubit shift operators can be written as
\[S^- = \sum\limits_{j=0}^{N-1} \ket{j}\bra{j+1}, \qquad S^+ = (S^-)^\dag =  \sum\limits_{j=0}^{N-1} \ket{j+1}\bra{j}.\]
Consider an integer $j$ with $0 \le j \le 2^{n-1}$. Its binary representation is denoted as $j = (j_{n-1}, \cdots , j_1, j_0)$ with the most significant bit on the left, where each $j_i \in \{0,1\}$ for $i = 0,1,\cdots,n-1$ and
\[j = j_{n-1} 2^{n-1} + \cdots  + j_1 2^1 + j_0 2^0.\]

For quantum circuits, we use the convention that the qubits in a circuit diagram are numbered increasingly from the top to the bottom as $0,1,\cdots,n-1$. An integer $j = (j_{n-1}, \cdots , j_1, j_0)$ input to a circuit is prepared as a set of quantum states $\ket{j_{n-1}}, \cdots, \ket{j_0}$ with $\ket{j_{n-1}}$ mapped to the lowest numbered qubit $q_0$ and $\ket{j_0}$ mapped to the highest numbered qubit $q_{n-1}$.  Note that this order contrasts with that used in Refs.~\cite{Sato24Circuit,HuJin24SchrCircuit}, aligning more closely with conventions typical in quantum computation. For instance, in a two-qubit system, we often use $\ket{01}$ and $\ket{10}$ to represent the computational basis states $\ket{1}$ and $\ket{2}$, respectively. This is because $(01) = 0 \times 2^1 + 1 \times 2^0 = 1$ and $(10) = 1 \times 2^1 + 0 \times 2^0 = 2$.

The decomposition of $S^-$ is based on the binary representation of $(j+1) - j = 1$, where $j = (j_{n-1}, \cdots , j_1, j_0)$. Noting that
$1 = (0, \cdots, 0, 1)$, we can enumerate all the possible pairs as follows:
\begin{alignat*}{3}
j & = (\cdots,  \cdots , 0),   & \qquad  j+1 & = (\cdots,  \cdots , 1), \\
j & = (\cdots,  \cdots, 0, 1),  & \qquad j+1  & = (\cdots,  \cdots , 1, 0), \\
j & = (\cdots,  \cdots, 0, 1,1),  &\qquad  j+1  & = (\cdots,  \cdots , 1, 0, 0), \\
& \hspace{2cm} \vdots    & \qquad &  \vdots    \\
j & = (0,  1,\cdots ,1),  & \qquad  j+1  & = (1,  0, \cdots , 0),
\end{alignat*}
where the missing binary digits are the same for $j$ and $j+1$. This implies
\begin{align*}
S^-
& = \sum_{j=1}^n I^{\otimes (n -j)} \otimes \Big( \ket{0,1,\cdots,1}  \bra{1, 0, \cdots, 0} \Big) \\
& = \sum_{j=1}^n I^{\otimes (n -j)} \otimes \ket{0}\bra{1} \otimes (\ket{1}\bra{0})^{\otimes (j-1)}.
\end{align*}
Therefore, we can represent the shift operators as
\begin{align*}
& S^- = \sum_{j=1}^n I^{\otimes (n -j)} \otimes \sigma_{01} \otimes \sigma_{10}^{\otimes (j-1)} =: \sum_{j=1}^n s_j^-, \\
& S^+ = \sum_{j=1}^n I^{\otimes (n -j)} \otimes \sigma_{10} \otimes \sigma_{01}^{\otimes (j-1)} =: \sum_{j=1}^n s_j^+.
\end{align*}

Let
\begin{align*}
S
& = \sigma_{01} \otimes \sigma_{10}^{\otimes (j-1)} + \sigma_{10} \otimes \sigma_{01}^{\otimes (j-1)} \\
& = \ket{0,1,\cdots,1}  \bra{1, 0, \cdots, 0} + \ket{1,0,\cdots,0}  \bra{0, 1, \cdots, 1}
= \ket{a_j}\bra{b_j} + \ket{b_j}\bra{a_j}
\end{align*}
be a $j$-qubit matrix, with
\[\ket{a_j} = \ket{0} \ket{1}^{\otimes(j-1)}, \qquad \ket{b_j} = \ket{1} \ket{0}^{\otimes(j-1)} . \]
The construction of the quantum circuits in \cite{Sato24Circuit,HuJin24SchrCircuit} is based on the following diagonalisation of $S$.
\begin{lemma}\label{lem:diagS}
The eigenvalues of $S$ are $\pm 1, \underbrace{0, \cdots, 0}_{2^j-2}$, and there exists a unitary matrix $B$ such that
\[S = B \left( Z \otimes (\ket{1}\bra{1})^{\otimes (j-1)} \right) B^\dag,\]
 where $Z = \text{diag}(1,-1)$ is the Pauli-$Z$ matrix and $B$ can be constructed by
\begin{equation}\label{operB}
\hat{B}: = \hat{H} \prod_{m=1}^{j-1} \text{CNOT}_m^0  ,
\end{equation}
where $\hat{H}$ is the Hadamard gate acting on the first qubit (i.e., qubit 0) and $\text{CNOT}_m^0$ is the CNOT gate acting on the $m$-th qubit controlled by the first qubit, with the index $m$ starting from 0.
\end{lemma}

It should be noted that we do not adhere to conventional operator actions, as in quantum circuits unitary operators are typically treated as quantum gates and operate via tensor products. For example, the unitary operators in \eqref{operB} are interpreted as $j$-qubits matrices, with the action considered to be the product between matrices (Note that $\prod$ is not the tensor product).

\begin{proof}
The matrix $S$ and the vectors $\ket{a_j}$ and $\ket{b_j}$ can be explicitly written as
\[S = \left[\begin{array}{ccc:ccc}
0   &         &     &      &      &     \\
    & \ddots  &     &      &      &     \\
    &         &  0  &  1   &      &     \\
\hdashline
    &         &  1  &  0    &      &     \\
    &         &     &       &  \ddots   &     \\
    &         &     &       &      &  0
\end{array}\right]_{2^j \times 2^j}, \qquad
\ket{a_j} = \left[\begin{array}{c}
0 \\
\vdots \\
1 \\
\hdashline
0 \\
\vdots \\
0
\end{array}\right] = \ket{2^{j-1}-1},\qquad
\ket{b_j} = \left[\begin{array}{c}
0 \\
\vdots \\
0 \\
\hdashline
1 \\
\vdots \\
0
\end{array}\right]= \ket{2^{j-1}},\]
where $\ket{2^{j-1}-1}$ and $\ket{2^{j-1}}$ are the computational basis states for a $j$-qubit system.  From $\braket{a_j}{b_j} = 0$ we have $S \ket{a_j} = \ket{b_j}$  and $S \ket{b_j} = \ket{a_j}$, leading to
\begin{align}
& S \frac{\ket{a_j} + \ket{b_j}}{\sqrt{2}} = \frac{\ket{a_j} + \ket{b_j}}{\sqrt{2}} = \frac{1}{\sqrt{2}} [0,\cdots, 1, 1, \cdots, 0]^T, \label{S1}\\
& S \frac{\ket{a_j} - \ket{b_j}}{\sqrt{2}} = - \frac{\ket{a_j} - \ket{b_j}}{\sqrt{2}} = - \frac{1}{\sqrt{2}} [0,\cdots, 1, -1, \cdots, 0]^T.\label{S2}
\end{align}

Let $\Lambda = Z \otimes (\ket{1}\bra{1})^{\otimes (j-1)} = \text{diag}(0,\cdots,0,1, 0, \cdots, 0, -1)$ be the diagonalization matrix for $S$. Then, one can choose a unitary matrix $B$ such that $S = B\Lambda B^\dag$, with
\[B = \frac{1}{\sqrt{2}} \left[\begin{array}{ccc:ccc}
1   &         &       &   1  &      &          \\
    & \ddots  &       &      &  \ddots    &      \\
    &         &   1   &     &      &  1     \\
\hdashline
    &         &   1   &      &      &   -1\\
    &  \begin{sideways}$\ddots$\end{sideways} &      &    &   \begin{sideways}$\ddots$\end{sideways}   &   \\
 1  &         &       & -1     &      &
\end{array}\right],\]
where the last columns in the left half and right half correspond to \eqref{S1} and \eqref{S2}, respectively. The additional columns are included while ensuring orthogonality.

One can check that the matrix $B$ can be constructed by
\[ \hat{B}: = \hat{H} \prod_{m=1}^{j-1} \text{CNOT}_m^0  ,\]
where $\hat{H}$ is the Hadamard gate acting on the first qubit (i.e., qubit 0) and $\text{CNOT}_m^0$ is the CNOT gate acting on the $m$-th qubit controlled by the first qubit. This can be verified by
\[B\ket{l} = \hat{B}\ket{l}, \qquad l = 0,1,\cdots, 2^j-1,\]
where $B\ket{l}$ is the $l$-th column of $B$, with the column index starting from 0.
\begin{itemize}
  \item If $l = (0, k_1,\cdots,k_{j-1})$, then $l \le 2^{j-1}-1$. This means $B\ket{l}$ is the column vector in the left half of $B$. Noting that
    \[2^j = (0, k_1,\cdots,k_{j-1}) + (1, 1-k_1,\cdots,1-k_{j-1} ),\]
  we have
  \begin{align*}
  B\ket{l}
  & = \frac{1}{\sqrt{2}} \Big( \ket{0, k_1,\cdots,k_{j-1} } + \ket{1, 1-k_1,\cdots,1-k_{j-1} } \Big) \\
  & = \hat{H} \text{CNOT}_1^0  \cdots \text{CNOT}_{j-1}^0 \ket{0, k_1,\cdots,k_{j-1} }.
  \end{align*}

  \item If $l = (1, k_1,\cdots,k_{j-1})$, then $l \ge 2^{j-1}$. This means $B\ket{l}$ is the column vector in the right half of $B$. And we have
  \begin{align*}
  B\ket{l}
  & = \frac{1}{\sqrt{2}} \Big( \ket{0, k_1,\cdots,k_{j-1} } - \ket{1, 1-k_1,\cdots,1-k_{j-1} } \Big) \\
  & = \hat{H} \text{CNOT}_1^0  \cdots \text{CNOT}_{j-1}^0 \ket{1, k_1,\cdots,k_{j-1} }.
  \end{align*}
\end{itemize}
The proof is completed.
\end{proof}

Similar result in Lemma \ref{lem:diagS} holds for
\begin{align*}
S(\lambda)  = \e^{\i \lambda} \sigma_{01} \otimes \sigma_{10}^{\otimes (j-1)} + \e^{-\i \lambda}  \sigma_{10} \otimes \sigma_{01}^{\otimes (j-1)},
\end{align*}
with the matrix $B$ constructed by
\[
\hat{B}_j(\lambda) = P(-\lambda) \hat{H} \prod_{m=1}^{j-1} \text{CNOT}_m^0,
\]
where $P(\lambda) = \text{diag}(1, \e^{\i\lambda})$ is the phase gate acting on the qubit 0.

\section{Quantum circuit for the Schr\"odingerised system} \label{sec:SchrLap}

In the context of the discrete heat equation, the coefficient matrix \( A \) associated with \eqref{ODElinear} is symmetric, leading to $ H_1 = A $ and $H_2 = O$ for the Schr\"odingerised system \eqref{generalSchr}. Given the similarity in structure of the coefficient matrices for the three-type boundary conditions, this section primarily focuses on the construction of the quantum circuit under the Dirichlet boundary condition.

As in \cite{HuJin24SchrCircuit}, the domain of $p$ is now chosen as $[-\pi R, \pi R]$, where $R$ is a positive number satisfying $\e^{-\pi R} \approx 0$. The Schr\"odingerised Laplacian in $d$ dimensions for the Dirichlet boundary condition is defined by
\begin{equation}\label{Hheat}
\mathbf{H}_{\text{heat}} = a \bb{D}_D^\Delta \otimes D_\eta,
\end{equation}
where $a$ is a constant, $D_\eta = \text{diag}(\eta_0, \cdots, \eta_{N_p-1})$ and $\eta_k = (k - \frac{N_p}{2})/R$ and $D_D^\Delta$ in \eqref{heatmatrixd} is an $n_x$-qubit matrix. In what follows, we briefly review the construction of the quantum circuit for the Hamiltonian evolution operator $ U_{\text{heat}}(\tau) := \e^{\i \mathbf{H}_{\text{heat}} \tau}$, as proposed in \cite{HuJin24SchrCircuit}, and discuss some modifications to both the gate construction and the time complexity.

\subsection{The approximate time evolution operator}

The evolution operator $ U_{\text{heat}}(\tau) = \e^{\i \mathbf{H}_{\text{heat}} \tau}$  can be formulated as
\begin{equation}\label{Uheat}
U_{\text{heat}}(\tau)
= \sum_{k=0}^{N_p-1} \left( \prod_{\alpha=1}^{d} \left(U_0(\tau)\right)_{\alpha} \right)^{k-N_p/2} \otimes |k\rangle \langle k|,
\end{equation}
where
\begin{equation}\label{U0tau}
U_0(\tau)= \e^{\i\mathbf{H}_{0}\tau}, \qquad
    \mathbf{H}_{0} := \gamma_{0} \left[ \sum_{j=1}^{n_x} ( s_{j}^{-} +  s_{j}^{+}) - 2 I^{\otimes n_x} \right], \quad  \gamma_{0} = \frac{a}{\Delta x^2 R}.
\end{equation}

The operator $U_0(\tau) = \e^{\i\mathbf{H}_{0}\tau}$ is approximated by
\begin{equation}\label{V0tau}
V_0(\tau) = \text{Ph}(-2\gamma_{0}\tau) \prod_{j=1}^{n_x} I^{\otimes(n_x-j)} \otimes W_j(\gamma_{0}\tau, 0),
\end{equation}
where $\text{Ph}(\theta)=\e^{i\theta} I^{\otimes n_x}$ is the global phase gate, so it doesn't affect the total evolution.  For simplicity, we label the qubits for $W_j(\gamma_{0}\tau, 0)$ from left to right as $0,1,\cdots, j-1$ for fixed $j$. That is, we only focus on the register given by the $j$ qubits on the right-hand side. The operator
\[W_{j}(\gamma\tau, \lambda) = B_{j}(\lambda) \e^{i\gamma\tau Z \otimes (|1\rangle\langle 1|)^{\otimes(j-1)}}  B_{j}(\lambda)^{\dagger}, \]
where
\[\e^{\i\gamma\tau Z \otimes (|1\rangle\langle 1|)^{\otimes(j-1)}} = \text{CRZ}_{j}^{1,\dots,j-1}(-2\gamma\tau)\]
is the multi-controled RZ gate acting on the $j$-th qubit controlled by $1,\dots,(j-1)$-th qubits. A detailed construction of this gate is given below. According to the definition of $Z$,
\[Z \otimes (|1\rangle\langle 1|)^{\otimes(j-1)}
= (\ket{0}\bra{0}  - \ket{1}\bra{1})\otimes (|1\rangle\langle 1|)^{\otimes(j-1)}\]
is a diagonal matrix, which gives
\begin{align*}
\e^{\i\gamma\tau Z \otimes (|1\rangle\langle 1|)^{\otimes(j-1)}}
& = (\ket{0}\bra{0}\e^{\i\gamma\tau}  - \ket{1}\bra{1}\e^{-\i\gamma\tau})\otimes (|1\rangle\langle 1|)^{\otimes(j-1)}
+ \sum_{k_1,\cdots,k_j} \ket{k_1,\cdots,k_j}\bra{k_1,\cdots,k_j}\\
& = R_z(-2 \gamma \tau ) \otimes (|1\rangle\langle 1|)^{\otimes(j-1)}
+ \sum_{k_1,\cdots,k_j} \ket{k_1,\cdots,k_j}\bra{k_1,\cdots,k_j}
\end{align*}
where $(k_1,\cdots,k_j)\ne (0,1,\cdots,1)$ and $(1,1,\cdots,1)$, and
\begin{equation}\label{Rz}
R_z(\theta)\equiv \e^{-\i \theta Z/2}
     = \cos \frac{\theta}{2} I - \i \sin \frac{\theta}{2} Z
     = \begin{bmatrix}
        \e^{-\i \theta /2}       &  0 \\
             0                   &  \e^{\i \theta /2}
      \end{bmatrix}.
\end{equation}
Obviously,
\begin{align*}
\sum_{k_1,\cdots,k_j} \ket{k_1,\cdots,k_j}\bra{k_1,\cdots,k_j}
& = I^{\otimes n} - \ket{0,1,\cdots,1}\bra{0,1,\cdots,1} - \ket{1,1,\cdots,1}\bra{1,1,\cdots,1} \\
& = I^{\otimes n} - \Big( (\ket{0}\bra{0}  + \ket{1}\bra{1})\otimes (|1\rangle\langle 1|)^{\otimes(j-1)}   \Big)\\
& = I^{\otimes n} - \Big( I \otimes (|1\rangle\langle 1|)^{\otimes(j-1)}   \Big)
 = I \otimes (I^{\otimes(j-1)} - (|1\rangle\langle 1|)^{\otimes(j-1)}).
\end{align*}
The above calculation shows that
\[\text{CRZ}_{j}^{1,\dots,j-1}(\theta) = R_z(\theta) \otimes (|1\rangle\langle 1|)^{\otimes(j-1)} + I \otimes (I^{\otimes(j-1)} - (|1\rangle\langle 1|)^{\otimes(j-1)}),\]
which is exactly the multi-controled RZ gate acting on the $j$-th qubit controlled by $1,\cdots,(j-1)$-th qubits.

Let $\tilde{V}_0(\tau) = \prod_{\alpha=1}^{d} \left(V_0(\tau)\right)_{\alpha}$. The Hamiltonian evolution $U_{\text{heat}}(\tau)$ is then approximated as
\begin{align}
U_{\text{heat}}(\tau)
& \approx V_{\text{heat}}(\tau) = \left( \tilde{V}^{-N_p/2}_{0}(\tau) \otimes I^{\otimes n_p} \right) \sum_{k=0}^{N_p-1} \tilde{V}_0^{k}(\tau) \otimes |k\rangle \langle k| \label{Vheat}\\
& =:\left( \tilde{V}^{-N_p/2}_{0}(\tau) \otimes I^{\otimes n_p} \right)  \text{SEL}(\tilde{V}_0(\tau)),\nonumber
\end{align}
where the summation is a select oracle frequently encountered in the LCU. This oracle is defined below.

\begin{definition}\cite{Childs2022Notes,Lin2022Notes}
Let $U_0,\cdots,U_{J-1}$ be unitary matrices with $J = 2^a$. Then a select oracle is defined by
\[U:= \sum_{j=0}^{J-1} \ket{j}\bra{j} \otimes U_j\qquad \mbox{or} \qquad \sum_{j=0}^{J-1} U_j \otimes \ket{j}\bra{j}\]
which implements the selection of $U_i$ conditioned on the value of the $a$-qubit ancilla states (also called the control register).
\end{definition}

\begin{lemma}\label{lem:Vj}
If $U_j = V^j$, where $V$ is a unitary matrix, then the select oracle $U = \sum_{j=0}^{J-1} U_j \otimes \ket{j}\bra{j}$ can be realized by performing the controlled operator $c-V^{2^m}$ on the first register,  controlled by the $m$-th qubit of the second register if it is 1.
\end{lemma}
\begin{proof}
If the second register is $\ket{j}$, with the binary representation given by $j = (j_{a-1}, \cdots, j_0) = j_{a-1}2^{a-1}+ \cdots + j_02^0$, then the operation on the first register is actually
\[V^{j_{a-1}2^{a-1}} \cdots  V^{j_02^0} = V^{j_{a-1}2^{a-1}+ \cdots + j_02^0} = V^j,\]
where $V^{j_l2^l} = I$ is naturally included if $j_l = 0$.
\end{proof}

In the LCU approach, we need the select oracle for the unitaries $U_j$ to be added together. This can be constructed with only $\log J $ queries to controlled $U_j$'s if each $U_j$ can be efficiently implemented as will be observed in Section \ref{subsec:modification}.

\begin{figure}[!htb]
  \centering
  \includegraphics[scale=0.2]{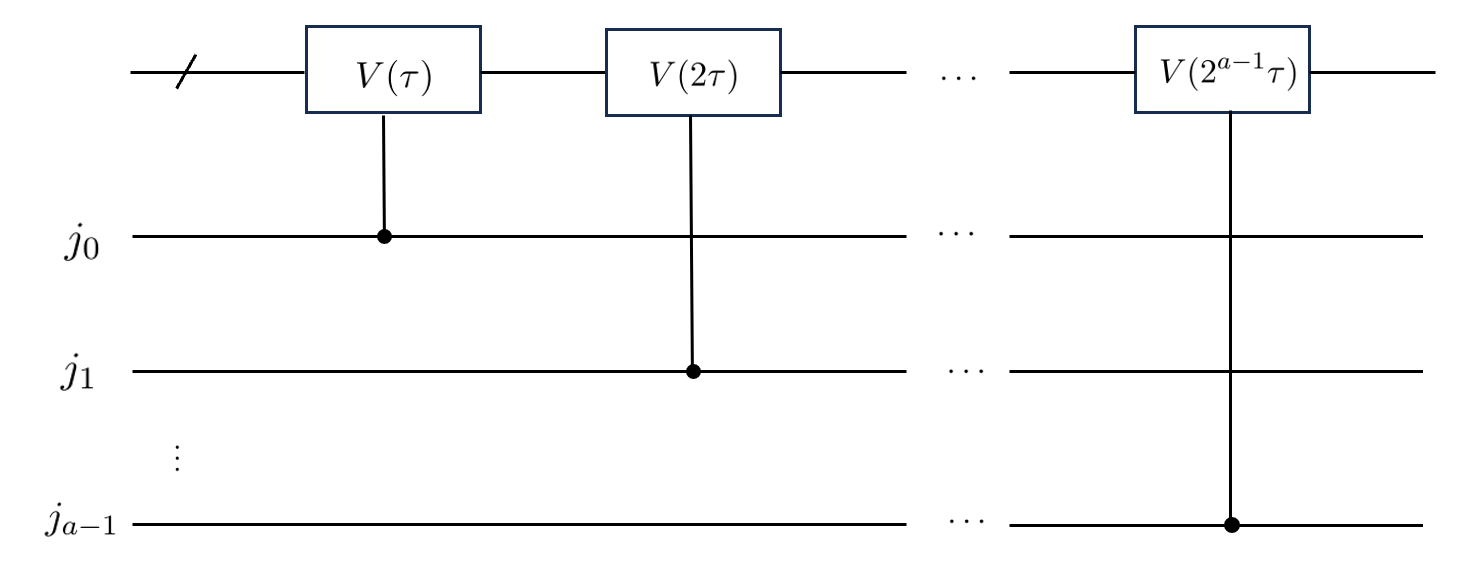}\\
  \caption{Quantum circuit for the select oracle given in Lemma \ref{lem:selectH}}\label{fig:selectV}
\end{figure}

\begin{lemma}\label{lem:selectH}
 Let $U_j(\tau)$ be a set of unitaries such that $U_j(\tau)$ = $V^j(\tau) = V(j\tau)$. Assume that the complexity of $V(s)$ is independent of the parameter $s$. Then the select oracle $U = \sum_{j=0}^{J-1} U_j \otimes \ket{j}\bra{j}$ can be constructed with $\log J$ queries to $V$.
\end{lemma}
\begin{proof}
According to Lemma \ref{lem:Vj}, it holds that
\[V^{2^l}(\tau) = V(2^l\tau ), \quad \text{for } l = 0, 1, \ldots, a-1, \quad a = \log J.\]
The quantum circuit illustrating this construction is depicted in Fig.~\ref{fig:selectV}. According to the assumption,  the number of queries to $V$ is $a = \log J$.
\end{proof}

\begin{remark}  \label{rem:logJ}
The requirement of $\log J$ queries to $V$ assumes that the complexity of $V(s)$ remains consistent regardless of the parameter $s$. It is crucial to note that this assumption does not hold for the time evolution operator $V(s) = \e^{\i \mathbf{H} s}$ in practical scenarios. Typically, we only have access to the evolution over a single segment and extrapolate using expected times to reach the final state. However, we can always assume access to $V(s) = \e^{\i \mathbf{H} s}$ for $s\le s_0$, where $s_0$ is a small duration. Then the select oracle $U = \sum_{j=0}^{J-1} U_j \otimes \ket{j}\bra{j}$ employs $\log J$ queries to $V$ if $Js \le s_0$. At this point, the time complexity will be reflected in the number of repetitions $T/s$.
\end{remark}

\begin{figure}[!htb]
  \centering
  \includegraphics[scale=0.2]{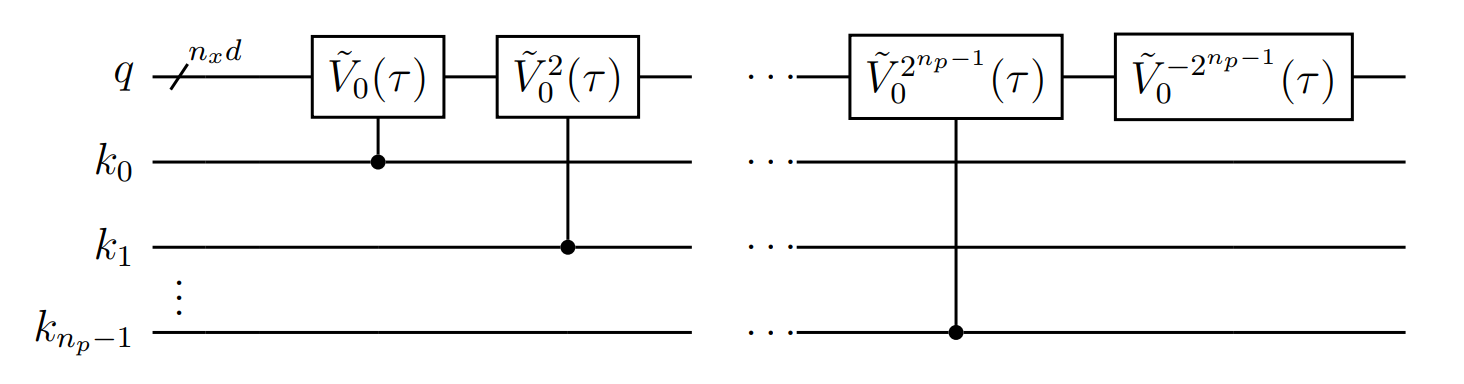}\\
  \caption{Quantum circuit for $V_{\text{heat}}(\tau)$}\label{fig:Vheat}
\end{figure}

\begin{figure}[!htb]
  \centering
  \includegraphics[scale=0.2]{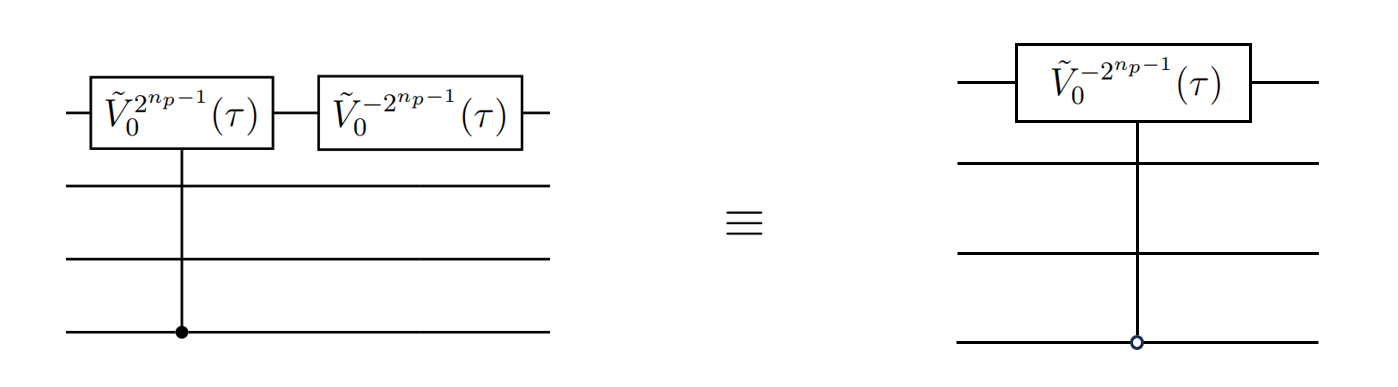}\\
  \caption{Modified quantum circuit for the last two operators in $V_{\text{heat}}(\tau)$}\label{fig:Vheatm}
\end{figure}

Since we have reversed the order of quantum states in the $x$ register in Refs.~\cite{Sato24Circuit, HuJin24SchrCircuit}, the parallel lines in the quantum circuits should similarly reverse direction from top to bottom in the $x$ register. The specific quantum circuits for $W_j(\gamma \tau, \lambda)$, $V_0(\tau)$, and $\tilde{V}_0(\tau)$ are not depicted here. Interested readers are directed to Figs.~1-3 in \cite{HuJin24SchrCircuit}. The approximate Hamiltonian evolution operator $V_{\text{heat}}(\tau)$ is shown in Fig.~\ref{fig:Vheat}. We remark that the last two operators in Fig.~\ref{fig:Vheat} is equivalent to the right one in Fig.~\ref{fig:Vheatm}, where the controlled gate is only active if $\ket{k_{n_p-1}}$ is in the 0 state.

\subsection{Modification to the select oracle} \label{subsec:modification}

In Lemma 3 of \cite{HuJin24SchrCircuit}, the complexity of the quantum circuit depicted in Fig.~\ref{fig:Vheat} is evaluated based on the count of controlled $\tilde{V}_0(\tau)$ gates, resulting in
\[
\sum_{m=0}^{n_p-1} 2^m = 2^{n_p}-1 = \mathcal{O}(N_p)
\]
uses of $\tilde{V}_0(\tau)$.
However, it's worth noting that this count significantly overestimates the actual number. The analysis leading to  this assertion is discussed below.

\begin{figure}[!htb]
  \centering
  \includegraphics[width=0.8\textwidth]{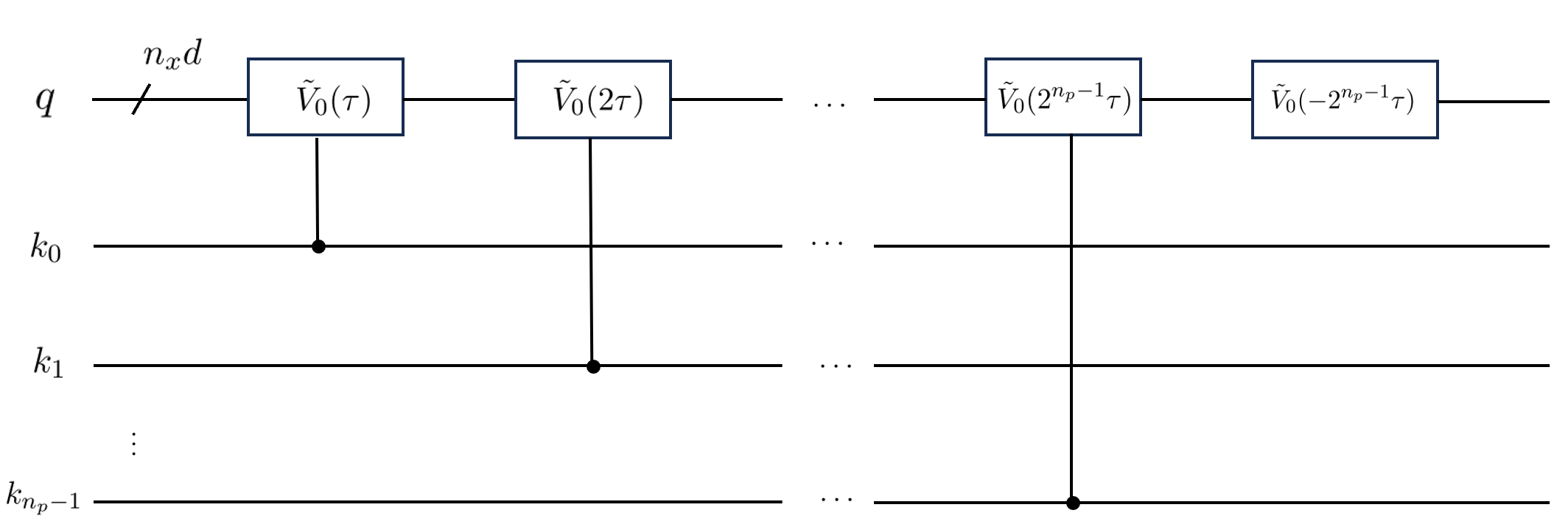}\\
  \caption{Modified quantum circuit for $V_{\text{heat}}(\tau)$}\label{fig:Vheatm1}
\end{figure}

The operator $\tilde{V}_0^{2^m}(\tau)$ in Fig.~\ref{fig:Vheat} is defined by
\[\tilde{V}_0^{2^m}(\tau) = \Big(\prod_{\alpha=1}^d (V_0(\tau))_\alpha\Big)^{2^m}, \qquad m = 0,1,\cdots, n_p-1,\]
where
\[(V_0(\tau))_\alpha = I^{\otimes (d-\alpha) n_x} \otimes V_0(\tau) \otimes I^{\otimes (\alpha-1) n_x},\]
which gives
\begin{align*}
\tilde{V}_0^{2^m}(\tau)
  = (V_0(\tau) \otimes \cdots \otimes V_0(\tau))^{2^m} = V_0^{2^m}(\tau) \otimes \cdots \otimes V_0^{2^m}(\tau)
  = \prod_{\alpha=1}^d (V_0(\tau)^{2^m})_\alpha.
\end{align*}
By the definition of $V_0(\tau)$ in \eqref{V0tau}, one has
\[V_0(\tau)^{2^m} = V_0(2^m \tau), \qquad m = 0,1,\cdots, n_p-1,\]
which immediately leads to
\[\tilde{V}_0^{2^m}(\tau) = \tilde{V}_0(2^m\tau), \qquad m = 0,1,\cdots, n_p-1.\]
Therefore, $\tilde{V}_0^{2^m}(\tau)$ in the quantum circuit can be replaced by $\tilde{V}_0(2^m \tau)$, as illustrated in Fig.~\ref{fig:Vheatm1}. The last two operators depicted can be consolidated into the single operator shown on the right in Fig.~\ref{fig:Vheatm}. {\it  The significant reduction in complexity within one segment arises from the fact that $\tilde{V}_0(2^m\tau)$ uses the same number of gates regardless of the parameter of $\tilde{V}_0(s)$.}

For the Dirichlet and Neumann boundary conditions, the $n_x$-qubit coefficient matrices are identical. Therefore, the time evolution can be directly executed using the approximate time evolution operator $V_{\text{heat}}(\tau)$, with the gate complexity described as follows.

\begin{lemma}\label{lem:Vheat}
 The approximated time evolution operator $V_{\text{heat}}(\tau)$ in \eqref{Vheat} can be implemented using $\mathcal{O}(d n_p n_x)$ single-qubit gates and at most $\mathcal{Q}_{V_{\text{heat}}} = \mathcal{O}(d n_p n_x^2)$ CNOT gates for $n_x \geq 3$.
\end{lemma}

\begin{proof}
Since $\tilde{V}_0^{2^m}(\tau) = \tilde{V}_0(2^m\tau)$ for $m = 0, 1, \ldots, n_p-1$, the operation of the modified quantum circuit in Fig.~\ref{fig:Vheatm1} effectively achieves the same outcome as that of the quantum circuit in Fig.~\ref{fig:Vheat}. Therefore, instead of $\mathcal{O}(2^{n_p})$ applications of controlled $\tilde{V}_0(\tau)$ gates, we only require $\mathcal{O}(n_p)$ applications of $\tilde{V}_0(2^m\tau)$ ($m=0,\cdots,n_p-1$) gates (see also Lemma \ref{lem:selectH}). Consequently, the parameter $N_p = 2^{n_p}$ can be replaced by the logarithmic factor $n_p = \log N_p$.
\end{proof}

\begin{remark}
Reducing the computational overhead from discretizing the auxiliary variable \( p \) in the Schr\"odingerisation method is possible. We outline three strategies to achieve this goal:

\begin{enumerate}
  \item Gate count reduction in one segment:
  The original circuit in \cite{HuJin24SchrCircuit} featured a factor \( N_p \) per segment, which has now been replaced by a logarithmic factor \( n_p \).

  \item Improving dependence on time steps:
  In \cite[Lemma 2]{HuJin24SchrCircuit}, the approximation error in terms of operator norm is bounded by
   \[ \left\| \e^{\i \mathbf{H} \tau } - V_{\text{heat}}(\tau) \right\| \le \frac{d N_p (n_x-1)\gamma_0^2}{4} \tau^2, \]
   indicating a linear dependence on \( N_p \) for repeated steps, due to the use of first order Trotter splitting in Eq. (3.18).  This can be improved to  \( N_p^{1/q} \) if employing \( q \)-th order Trotter splitting.

  \item Accuracy in \( p \) variable discretization: Achieving arbitrary-order accuracy in \( p \) is feasible by selecting smoother initial data for \eqref{heatww}, as demonstrated in \cite{JLM24SchrInhom,JLM24SchrBackward}. Here, \( (\Delta p)^l \sim \varepsilon \) for order \( l \) implies \( N_p \sim 2\pi R/\Delta p \sim 2\pi R \varepsilon^{-1/\ell} \) and \( n_p \sim \log( 2\pi R \varepsilon^{-1/\ell} ) \).
\end{enumerate}

\end{remark}

\subsection{Discussion on the periodic boundary condition}

For the periodic boundary condition, we need to deal with the additional term $\frac{1}{\Delta x^2} (\sigma_{01}^{\otimes n_x} + \sigma_{10}^{\otimes n_x})$ in \eqref{DPDelta}. The Schr\"odingerised Laplacian in $d$ dimensions in this case is defined by
\begin{equation}\label{HheatP}
\mathbf{H}_{\text{heat}}^P = a \bb{D}_P^\Delta \otimes D_\eta,
\end{equation}
where
\[
\bb{D}_P^\Delta = \sum_{\alpha=1}^d (D_P^\Delta)_\alpha,
\]
with
\begin{align*}
 D_P^\Delta = D_D^\Delta + \frac{1}{\Delta x^2} (\sigma_{01}^{\otimes n_x} + \sigma_{10}^{\otimes n_x}) ,\qquad
 (\bullet)_\alpha := I^{\otimes (d-\alpha) n_x} \otimes \bullet \otimes I^{\otimes (\alpha-1) n_x}.
\end{align*}
A direct manipulation gives
\begin{align*}
\mathbf{H}_{\text{heat}}^P
& = a \bb{D}_P^\Delta \otimes D_\eta =  a \sum_{\alpha=1}^d (D_P^\Delta )_\alpha \otimes D_\eta \\
& = \sum_{k=0}^{N_p-1} \frac{a \eta_k}{\Delta x^2} \sum_{\alpha=1}^d \Big(\sum_{j=1}^{n_x} (s_j^-+s_j^+) - 2 I^{\otimes n_x}  + (\sigma_{01}^{\otimes n_x} + \sigma_{10}^{\otimes n_x})\Big)_\alpha \otimes \ketbra{k}{k} \\
& = \sum_{k=0}^{N_p-1} \Big( k - \frac{N_p}{2}\Big) \sum_{\alpha=1}^d (\mathbf{H}_1)_\alpha \otimes \ketbra{k}{k},
\end{align*}
where
\[\mathbf{H}_1 = \gamma_0 \Big[ \sum_{j=1}^{n_x} (s_j^-+s_j^+) - 2 I^{\otimes n_x}  + (\sigma_{01}^{\otimes n_x} + \sigma_{10}^{\otimes n_x})\Big] = \mathbf{H}_0 + \gamma_0(\sigma_{01}^{\otimes n_x} + \sigma_{10}^{\otimes n_x}), \qquad \gamma_0 = \frac{a}{\Delta x^2 R}.\]

By a direct manipulation, we can write the Hamiltonian evolution operator $U_{\text{heat}}^P(\tau) := \e^{\i \mathbf{H}_{\text{heat}}^P \tau}$ as follows,
\begin{align*}
U_{\text{heat}}^P(\tau)
& = \exp\Big( \i \tau \sum_{k=0}^{N_p-1} \Big( k - \frac{N_p}{2}\Big) \sum_{\alpha=1}^d (\mathbf{H}_1)_\alpha \otimes \ketbra{k}{k}  \Big) \\
& = \sum_{k=0}^{N_p-1} \Big(\e^{ \i \tau \sum_{\alpha=1}^d (\mathbf{H}_1)_\alpha }\Big)^{k-N_p/2}\otimes \ketbra{k}{k}
  = \sum_{k=0}^{N_p-1} \Big(\prod_{\alpha=1}^d\e^{ \i \tau (\mathbf{H}_1)_\alpha }\Big)^{k-N_p/2}\otimes \ketbra{k}{k} \\
& = \sum_{k=0}^{N_p-1} \Big(\prod_{\alpha=1}^d (U_1(\tau))_\alpha\Big)^{k-N_p/2}\otimes \ketbra{k}{k},
\end{align*}
where $U_1(\tau) := \e^{ \i \tau \mathbf{H}_1 }$ can be treated as $U_0(\tau)$ in \eqref{Uheat}. Therefore, the quantum circuit can be derived similarly to the case of the Dirichlet boundary condition when obtaining the quantum circuit of $U_1(\tau)$.

By employing the first-order Lie-Trotter-Suzuki decomposition, $U_1(\tau)$ is approximated as
\begin{align*}
U_1(\tau)
& = \exp\Big( \i \tau \gamma_0  \Big[ \sum_{j=1}^{n_x} (s_j^-+s_j^+) - 2 I^{\otimes n_x}  + (\sigma_{01}^{\otimes n_x} + \sigma_{10}^{\otimes n_x})\Big] \Big) \\
& \approx \e^{\i \mathbf{H}_0 \tau} \exp\Big(\i \tau \gamma_0 (\sigma_{01}^{\otimes n_x} + \sigma_{10}^{\otimes n_x})\Big) = U_0(\tau) U_1^{(1)}(\tau) = V_1(\tau),
\end{align*}
where $U_0(\tau) = \e^{\i \mathbf{H}_0 \tau}$ is defined in \eqref{U0tau} and can be approximated by $V_0(\tau)$, and
\[U_1^{(1)}(\tau) := \exp\Big(\i \tau \gamma_0 (\sigma_{01}^{\otimes n_x} + \sigma_{10}^{\otimes n_x})\Big).\]
Now, we can treat the approximate operator $V_1(\tau)$ as equivalent to $V_0(\tau)$ in \eqref{V0tau}.

The quantum circuit for $U_1^{(1)}(\tau)$ is depicted in Fig.~6 of \cite{HuJin24SchrCircuit}. In this circuit, the parameter $\gamma_1$ should be replaced with $\gamma_0$, and the direction of the parallel lines in the diagram should be reversed from top to bottom.
The diagonalization in the Bell basis is presented in Eq.~(4.15) there, which can be analyzed similarly to Lemma \ref{lem:diagS}, given that $\sigma_{01}^{\otimes n_x} + \sigma_{10}^{\otimes n_x}$ can be transformed into $S$ through certain permutations.

Let $V_{\text{heat}}^P(\tau)$ be the approximate Hamiltonian evolution operator of $U_{\text{heat}}^P(\tau)$. Following the similar arguments in Lemma 6 of \cite{HuJin24SchrCircuit}, we are ready to get the gate complexity for the heat equation with the periodic boundary condition.

\begin{lemma}\label{lem:VheatP}
 The approximated time evolution operator $V_{\text{heat}}^P(\tau)$ for the periodic boundary condition can be implemented using $\mathcal{O}(dn_p n_x)$ single-qubit gates and at most $\mathcal{Q}_{V_{\text{heat}}^P} = \mathcal{O}(d n_p n_x^2)$ CNOT gates for $n_x \geq 3$.
\end{lemma}

\begin{remark} \label{rem:errHom}
Hereafter, the quantum circuit $V_{\text{heat}}(\tau)$ representing $\e^{\i \mathbf{H} \tau }$ will be regarded as a black box, serving as an input model for the time evolution, where $\mathbf{H}$ denotes the Schr\"odingerised Laplacian. The approximation error in terms of operator norm is bounded by
\[ \|\e^{\i \mathbf{H} \tau }-V_{\text{heat}}(\tau)\| \le C_{\text{heat}} \tau^2,\]
where
\[C_{\text{heat}} = \frac{d N_p (n_x-1)\gamma_0^2}{4}, \qquad \gamma_0 = \frac{a}{\Delta x^2 R}\]
is for the Dirchlet boundary condition (see Lemma 2 of \cite{HuJin24SchrCircuit}). Similar upper bounds can be obtained for the Neumann and periodic boundary conditions with the details omitted.
\end{remark}

\section{Linear combination of states  for the inhomogeneous term} \label{sec:LCU}

The Duhamel's principle in \eqref{Duhamel} for the discrete heat equation is
\begin{equation}\label{DuhamelHeat}
\tilde{\bb{w}}(T) = \e^{\i \mathbf{H} T } \tilde{\bb{w}}(0) + \int_0^T \e^{\i (T-s) \mathbf{H} } \tilde{\bb{b}}(s) \d s =:
\tilde{\bb{w}}_H + \tilde{\bb{w}}_b,
\end{equation}
where $\mathbf{H}$ is the discrete Laplacian. Based on this, we can prepare both vectors as quantum states and then perform a linear combination of the states. The quantum circuit for the Schr\"odingerised Laplacian has already addressed the first part.
However, the second part is no longer an operator acting on a vector and thus requires further treatment.
The approach involves first discretizing the integral using numerical quadrature, followed by applying another linear combination of states. Specifically, we utilize first-order numerical quadrature with $K$ equidistant nodes to get
\begin{align*}
\int_0^T \e^{\i (T-s) \mathbf{H} } \tilde{\bb{b}}(s) \d s
 \approx \Delta s \sum_{j = 0}^{K-1}\e^{\i \mathbf{H}(T-s_j)}\tilde{\bb{b}}(s_j)
 = \Delta s  \e^{\i \mathbf{H}T} \sum_{j = 0}^{K-1}(\e^{-\i \mathbf{H}\Delta s })^j \tilde{\bb{b}}(s_j)
 =: \tilde{\bb{w}}_b^a ,
\end{align*}
where $s_j = j \Delta s$, $\Delta s = T/K$ and
\begin{equation}\label{wb}
\tilde{\bb{w}}_b^a  = \Delta s\, \e^{\i \mathbf{H}T} \sum_{j = 0}^{K-1}(\e^{-\i \mathbf{H}\Delta s })^j \tilde{\bb{b}}(s_j) ,
\end{equation}
which is an approximation of
\[\tilde{\bb{w}}_b :=  \int_0^T \e^{\i (T-s) \mathbf{H} } \tilde{\bb{b}}(s) \d s.\]
It should be pointed out that we do not utilize the step size $\tau$ for numerical quadrature here, thereby ensuring that our analysis remains independent of the previously mentioned construction for the discrete Laplacian.

According to Lemma 34 in \cite{An2022blockEncodingODE}, it suffices to choose
\begin{equation*}
K = \Theta\Big( \frac{T^2}{\varepsilon} \sup_{t\in [0,T]} (\|\mathbf{H}\| \|\tilde{\bb{b}}(t)\| + \| \frac{\d }{\d t}\tilde{\bb{b}}(t)\|) \Big),
\end{equation*}
where $\varepsilon$ is the upper bound of the quadrature error.

The solution $\tilde{\bb{w}}(T)$ is then represented as
\begin{equation}\label{wT}
\tilde{\bb{w}}(T) =  \tilde{\bb{w}}_H + \tilde{\bb{w}}_b \approx \tilde{\bb{w}}^a_H + \tilde{\bb{w}}^a_b = \tilde{\bb{w}}^a(T),
\end{equation}
where $\tilde{\bb{w}}^a_H = V_\text{heat} \tilde{\bb{w}}(0)$ is an approximation of $\tilde{\bb{w}}_H$ with $\e^{\i \mathbf{H}T}$ replaced by the quantum circuits for the homogeneous problem, where $V_\text{heat} = V^r_\text{heat}(\tau)$ ($r = T/\tau$).
For this reason, we can utilize the LCU technique to achieve the final state.

\subsection{Implementation of the inhomogeneous term}

In this section we provide the implementation of
\[\tilde{\bb{w}}_b^a = \Delta s \e^{\i \mathbf{H} T } \sum_{j = 0}^{K-1}(\e^{-\i \mathbf{H}\Delta s })^j \tilde{\bb{b}}(s_j) \approx   \int_0^T \e^{\i (T-s) \mathbf{H} } \tilde{\bb{b}}(s) \d s = \tilde{\bb{w}}_b.\]
Let $\ket{\tilde{\bb{b}}(s_j)}$ be the quantum state of $\tilde{\bb{b}}(s_j)$. Then,
\[\tilde{\bb{w}}_b^a  = \Delta s U_K \sum_{j = 0}^{K-1}U_j \tilde{\bb{b}}(s_j)
=: U_K \sum_{j = 0}^{K-1} \alpha_j U_j \ket{\tilde{\bb{b}}(s_j)},\]
where
\[\alpha_j = \Delta s \|\tilde{\bb{b}}(s_j)\|,  \qquad U_j = (\e^{-\i \mathbf{H} \Delta s})^j,\]
and $\e^{-\i \mathbf{H} \Delta s}$ is exactly the Hamiltonian evolution operator with the quantum circuit given before.

In the following we present the details on implementing the LCU \cite{Lin2022Notes}. To this end, we should give the following oracles:
\begin{itemize}
  \item The coefficient oracle
  \[O_{\text{coef}}: \ket{0^{n_s}} \to \frac{1}{\sqrt{\|\alpha\|_1}} \sum\limits_{j=0}^{K-1}\sqrt{\alpha_i} \ket{j}, \qquad K = 2^{n_s},\]
where $\alpha = (\alpha_0, \cdots, \alpha_{K-1})$  and
  \[\|\alpha\|_1 = \alpha_0 + \cdots + \alpha_{K-1} = \Delta s( \|\tilde{\bb{b}}(s_0)\| + \cdots + \|\tilde{\bb{b}}(s_{K-1})\| ).\]

  \item  The one-segment time or Hamiltonian evolution input oracle
  \[O_{\mathbf{H}}  = \e^{\pm \i \mathbf{H} \Delta s} \qquad \mbox{or} \qquad \e^{\pm \i \mathbf{H} \tau},\]
  where $\Delta s$ is the step size for the numerical integration and $\tau$ is the time step for the homogeneous problems.

  \item The select oracle
  \[\text{SEL}_{\mathbf{H}} = \sum_{j=0}^{K-1} \ket{j}\bra{j} \otimes U_j, \qquad U_j = (\e^{-\i \mathbf{H} \Delta s})^j.\]

  \item The source term input oracle
\begin{equation}\label{Obsourceinput}
  O_b: \ket{j}\ket{0^{n_{xp}}} \to \ket{j}\ket{\tilde{\bb{b}}(s_j)}, \qquad n_{xp} = d n_x + n_p.
\end{equation}
  \item The state preparation oracle $O_{\text{prep}}: \ket{0^{n_{xp}}} \to \ket{\tilde{\bb{w}}(0)}$.
\end{itemize}

We now discuss the quantum implementation and its complexity for the inhomogeneous term, following standard arguments as in \cite{{ALL2023LCH}}.

\begin{lemma}\label{lem:inhomo}
There exists a quantum algorithm which maps $\ket{0^{n_s}} \otimes \ket{0^{n_{xp}}}$ to the state $\frac{1}{\|\alpha\|_1} \ket{0^{n_s}} \otimes \tilde{\bb{w}}_b^{aa} + \ket{\bot}$ such that $\tilde{\bb{w}}_b^{aa}$ is a $\delta_1$-approximation of $\tilde{\bb{w}}_b = \int_0^T \e^{\i (T-s) \mathbf{H} } \tilde{\bb{b}}(s)\d s$, using queries to the Hamiltonian evolution input oracle $O_{\mathbf{H}}$ a total number of times $\mathcal{O}(K)$, queries to the coefficient oracle $O_{\text{coef}}$ and the source term input oracle $O_b$ for $\mathcal{O}(1)$ times, where $K$ is given by
\[ K = \Theta \Big( \frac{T^2}{\delta_1 } \max\Big\{ 2 C_{\text{heat}} \|\tilde{\bb{b}}\|_{avg},\quad  \sup_{t \in [0,T]} (\|\mathbf{H}\| \|\tilde{\bb{b}}(t)\| + \| \frac{\d}{\d t}\tilde{\bb{b}}(t)\|)  \Big\}\Big).\]
\end{lemma}

\begin{proof}
In the implementation, the operator $\e^{-\i \mathbf{H} \Delta s}$ is approximated by the quantum circuit of $V_{\text{heat}}(\Delta s)$ in Section \ref{sec:SchrLap}. For this reason, we use $U_j^a$ to denote the approximation of $U_j$.
The procedure is as follows:
    \begin{align*}
\ket{0^{n_s}} \otimes \ket{0^{n_{xp}}}
&  \quad \xrightarrow{ O_{\text{coef}} \otimes I^{\otimes n_{xp}} } \quad \frac{1}{\sqrt{\|\alpha\|_1}} \sum\limits_{j=0}^{K-1}\sqrt{\alpha_i} \ket{j} \otimes \ket{0^{n_{xp}}} \\
&  \quad \xrightarrow{~~~~O_b ~~~~~~~~} \quad \frac{1}{\sqrt{\|\alpha\|_1}} \sum\limits_{j=0}^{K-1}\sqrt{\alpha_i} \ket{j} \otimes \ket{\tilde{\bb{b}}(s_j)}\\
&  \quad \xrightarrow{ ~~~\text{SEL}_{\mathbf{H}} ~~~~~} \quad \frac{1}{\sqrt{\|\alpha\|_1}} \sum\limits_{j=0}^{K-1}\sqrt{\alpha_i} \ket{j} \otimes U_j^a \ket{\tilde{\bb{b}}(s_j)} \\
&  \quad \xrightarrow{~O_{\text{coef}}^\dag \otimes I^{\otimes n_{xp}}} \quad \frac{1}{\|\alpha\|_1} \ket{0^{n_s}} \otimes \sum\limits_{j=0}^{K-1} \alpha_i U_j^a \ket{\tilde{\bb{b}}(s_j)} + \ket{\bot} \\
& \quad \xrightarrow{~~ I^{\otimes n_s} \otimes  U_K^a~~~} \quad  \frac{1}{\|\alpha\|_1} \ket{0^{n_s}} \otimes \tilde{\bb{w}}_b^{aa} + \ket{\bot'},
\end{align*}
where $\tilde{\bb{w}}_b^{aa}$ is the approximation of $\tilde{\bb{w}}_b^a$, and the fourth step is derived from
\[O_{\text{coef}}^\dag = \frac{1}{\sqrt{\|\alpha\|_1}}
\begin{bmatrix} \sqrt{\alpha_0}  &  \cdots   &  \sqrt{\alpha_{K-1}} \\
 *  & \cdots & * \\
 \vdots  & \ddots  & \vdots \\
  *  & \cdots & *
  \end{bmatrix}.\]
The error can be bounded by
\begin{align*}
 \|\tilde{\bb{w}}_b^{aa} - \tilde{\bb{w}}_b  \|
 \le \| \tilde{\bb{w}}_b^{aa} - \tilde{\bb{w}}_b^a\| +
 \|\tilde{\bb{w}}_b^a - \tilde{\bb{w}}_b\|.
\end{align*}

The error for the first term on the right-hand side arises from the quantum circuit approximation of $\e^{-\i \mathbf{H} \Delta s}$. One has
\begin{align*}
\| \tilde{\bb{w}}_b^{aa} - \tilde{\bb{w}}_b^a \|
& = \Big\| \Delta s (V_{\text{heat}}(\Delta s))^K \sum_{j = 0}^{K-1} (V_{\text{heat}}(\Delta s))^j \tilde{\bb{b}}(s_j)  - \Delta s (\e^{-\i \mathbf{H}\Delta s })^K \sum_{j = 0}^{K-1} (\e^{-\i \mathbf{H}\Delta s })^j  \tilde{\bb{b}}(s_j) \Big\| \\
& \le \Big\| \Delta s \sum_{j = 0}^{K-1}\Big( (V_{\text{heat}}(\Delta s))^j - (\e^{-\i \mathbf{H}\Delta s })^j \Big) \tilde{\bb{b}}(s_j) \Big\| \\
& \qquad + \Big\|\Delta s ( (\e^{-\i \mathbf{H}\Delta s })^K - (V_{\text{heat}}(\Delta s))^K) \sum_{j = 0}^{K-1} (\e^{-\i \mathbf{H}\Delta s })^j  \tilde{\bb{b}}(s_j) \Big\|\\
& \le \Delta s \| V_{\text{heat}}(\Delta s) - \e^{-\i \mathbf{H}\Delta s }\| \cdot \sum_{j = 0}^K j \|\tilde{\bb{b}}(s_j)\|
 \le K \| V_{\text{heat}}(\Delta s) - \e^{-\i \mathbf{H}\Delta s }\| \|\tilde{\bb{b}}\|_{avg},
 \end{align*}
where
\[\|\tilde{\bb{b}}\|_{avg} = \Delta s( \|\tilde{\bb{b}}(s_0)\| + \cdots + \|\tilde{\bb{b}}(s_{K-1})\| ).\]
In view of Remark \ref{rem:errHom}, there holds
\[\| V_{\text{heat}}(\Delta s) - \e^{-\i \mathbf{H}\Delta s }\| \le  C_{\text{heat}}\Delta s^2 = C_{\text{heat}}T^2/K^2,\]
giving
\begin{equation}\label{wbaaErr}
\| \tilde{\bb{w}}_b^{aa} - \tilde{\bb{w}}_b^a \| \le C_{\text{heat}}T^2  \|\tilde{\bb{b}}\|_{avg}/K.
\end{equation}
To get a $\delta_1$-approximation, we should choose $K$ such that $\| \tilde{\bb{w}}_b^{aa} - \tilde{\bb{w}}_b^a \| \le \delta_1/2$, yielding
\[K \ge 2 C_{\text{heat}}T^2 \|\tilde{\bb{b}}\|_{avg} /\delta_1.\]
The second term on the right-hand side is the numerical quadrature error. According to Lemma 34 in \cite{An2022blockEncodingODE}, the error satisfies
\[
\|\tilde{\bb{w}}_b^a - \tilde{\bb{w}}_b\| \le \frac{T^2}{2K} \sup_{t \in [0,T]} (\|\mathbf{H}\| \|\tilde{\bb{b}}(t)\| + \| \frac{\d}{\d t}\tilde{\bb{b}}(t)\|) \le \frac{\delta_1}{2},
\]
suggesting that we should choose
\[
K \ge \frac{T^2}{\delta_1 } \sup_{t \in [0,T]} (\|\mathbf{H}\| \|\tilde{\bb{b}}(t)\| + \| \frac{\d}{\d t}\tilde{\bb{b}}(t)\|).
\]
Therefore, we must choose $K$ such that
\[ K \ge \frac{T^2}{\delta_1 } \max\Big\{ 2 C_{\text{heat}} \|\tilde{\bb{b}}\|_{avg},\quad  \sup_{t \in [0,T]} (\|\mathbf{H}\| \|\tilde{\bb{b}}(t)\| + \| \frac{\d}{\d t}\tilde{\bb{b}}(t)\|)  \Big\}.\]

The complexity primarily hinges on the selection oracle $\text{SEL}_{\mathbf{H}}$ and the evolution operator $U_K^a$.  It is important to note that we assume access only to the evolution operator within a single segment, i.e., \( O_{\mathbf{H}}(s) = \exp(-\i \mathbf{H} \Delta s) \). Contrary to the $\log K$ queries in Lemma \ref{lem:selectH}, the actual number of queries is
\[ 1 + 2 + 2^2 + \cdots + 2^{n_s-1} + K =  2^{n_s}-1 + K = 2K-1 . \]
This completes the proof.
\end{proof}

\subsection{Linear combination of homogeneous and inhomogeneous terms}

Following the standard LCU procedure, we are ready to combine the two terms on the right-hand side of \eqref{wT}.

Using the state preparation oracle and the quantum circuits for the homogeneous problem, we obtain
\begin{align*}
\ket{0^{n_s}} \ket{0^{n_{xp}}}
& \quad \xrightarrow{ I^{\otimes n_s} \otimes O_{\text{prep}} } \quad  \frac{1}{\eta_0} \ket{0^{n_s}} \otimes \tilde{\bb{w}}(0)\\
& \quad \xrightarrow{ I^{\otimes n_s} \otimes V_{\text{heat}} } \quad  \frac{1}{\eta_0} \ket{0^{n_s}} \otimes \tilde{\bb{w}}_H^a,
\end{align*}
where $\eta_0 = \|\tilde{\bb{w}}(0)\|$. For brevity, we denote $V_0$ to be the combined unitary.
According to Lemma \ref{lem:inhomo}, there exists a unitary $V_1$ such that
\[\ket{0^{n_s}} \ket{0^{n_{xp}}} \quad \xrightarrow{ V_1 } \quad  \frac{1}{\eta_1} \ket{0^{n_s}} \otimes \tilde{\bb{w}}_b^{aa} + \ket{\bot},\]
where
\[\eta_1 = \|\alpha\|_1 = \Delta s( \|\tilde{\bb{b}}(s_0)\| + \cdots + \|\tilde{\bb{b}}(s_{K-1})\| ) =: \|\tilde{\bb{b}}\|_{avg}.\]

Let $R_t$ be a single-qubit rotation such that
\[R_t\ket{0} = \frac{1}{\sqrt{\eta_0 + \eta_1}}( \sqrt{\eta_0} \ket{0} + \sqrt{\eta_1} \ket{1}).\]
The desired procedure is given as follows:
    \begin{align*}
\ket{0} \otimes \ket{0^{n_s}} \otimes \ket{0^{n_{xp}}}
&  \quad \xrightarrow{ R_t \otimes I^{\otimes n_s} \otimes I^{\otimes n_{xp}} }
\quad
    \frac{1}{\sqrt{\eta_0 + \eta_1}}( \sqrt{\eta_0} \ket{0} + \sqrt{\eta_1} \ket{1}) \otimes \ket{0^{n_s}} \otimes \ket{0^{n_{xp}}} \nonumber\\
&  \quad \xrightarrow{\ketbra{0}{0}\otimes V_0 + \ketbra{1}{1}\otimes V_1}
\quad
   \frac{1}{\sqrt{\eta_0} \sqrt{\eta_0 + \eta_1}}\ket{0} \otimes \ket{0^{n_s}} \otimes \tilde{\bb{w}}_H^a \nonumber\\
&  \hspace{4cm} +  \frac{1}{\sqrt{\eta_1} \sqrt{\eta_0 + \eta_1}}\ket{1} \otimes \ket{0^{n_s}} \otimes \tilde{\bb{w}}^{aa}_b + \ket{\bot_1}\nonumber\\
& \quad \xrightarrow{~R_t^\dag \otimes I^{\otimes n_s} \otimes I^{\otimes n_{xp}}~}
\quad
\frac{1}{\eta_0 + \eta_1}\ket{0}\otimes \ket{0^{n_s}}
\otimes ( \tilde{\bb{w}}_H^a + \tilde{\bb{w}}_b^{aa}) + \ket{\bot_2}  \nonumber \\
& \hspace{4cm} = \frac{1}{\eta_0 + \eta_1}\ket{0}\otimes \ket{0^{n_s}}
\otimes  \tilde{\bb{w}}^{aa}(T) + \ket{\bot_2} \\
& \quad \xrightarrow{I \otimes I^{\otimes n_s} \otimes I^{\otimes n_x} \otimes F_p^\dag}
\quad
 \frac{1}{\eta_0 + \eta_1}\ket{0}\otimes \ket{0^{n_s}}
\otimes  \bb{w}^{aa}(T) + \ket{\bot_3},
\end{align*}
where $F_p$ is the discrete Fourier transform in the $p$ direction.

\begin{lemma}\label{lem:map2w}
There exists a quantum algorithm which maps $\ket{0} \otimes \ket{0^{n_s}} \otimes \ket{0^{n_{xp}}}$ to the state $\frac{1}{\eta_0 + \eta_1}\ket{0}\otimes \ket{0^{n_s}} \otimes  \bb{w}^{aa}(T) + \ket{\bot}$ such that $\bb{w}^{aa}(T)$ is a $\delta_1$-approximation of $\bb{w}(T)$, where
\[\eta_0 = \|\tilde{\bb{w}}(0)\| = \|\bb{w}(0)\|, \qquad \eta_1 = \|\alpha\|_1 = \|\tilde{\bb{b}}\|_{avg},\]
using
\begin{enumerate}
  \item $\mathcal{O}(r + K)$ queries to the time evolution input oracle $O_{\mathbf{H}}$,
  where
  \[r = \Theta \Big( \frac{C_{\text{heat}} T^2\|\bb{w}(0)\|}{\delta_1}\Big), \qquad
  K = \Theta \Big( \frac{ T^2}{\delta_1} \max\Big\{ C_{\text{heat}} \|\tilde{\bb{b}}\|_{avg},~ \sup_{t \in [0,T]} (\|\mathbf{H}\| \|\tilde{\bb{b}}(t)\| + \| \frac{\d}{\d t}\tilde{\bb{b}}(t)\|)  \Big\}\Big),\]
  with
  \[\|\tilde{\bb{b}}\|_{avg} = \Delta s( \|\tilde{\bb{b}}(s_0)\| + \cdots + \|\tilde{\bb{b}}(s_{K-1})\| ),\]

  \item $\mathcal{O}(1)$ queries to the state preparation oracle $O_{\text{prep}}$, the source term input oracle $O_b$ and the coefficient oracle $O_{\text{coef}}$.
\end{enumerate}
\end{lemma}
\begin{proof}
Assume negligible error from the quantum Fourier transform. To achieve a $\delta_1$-approximation of $\bb{w}(T)$, it is sufficient to satisfy
\[\|\tilde{\bb{w}}(T) - \tilde{\bb{w}}^{aa}(T)\| \le \|\tilde{\bb{w}}_H - \tilde{\bb{w}}_H^a \| + \|\tilde{\bb{w}}_b - \tilde{\bb{w}}_b^{aa}\| =: \varepsilon_0 + \varepsilon_1 \le \delta_1.\]

For $\varepsilon_0$, one can require
\begin{align*}
\varepsilon_0
 = \|\tilde{\bb{w}}_H - \tilde{\bb{w}}_H^a \| \le \|\e^{\i \mathbf{H} T } - V_{\text{heat}}\|  \|\tilde{\bb{w}}(0)\|
 = \|\e^{\i \mathbf{H} T } - V_{\text{heat}}\|  \|\bb{w}(0)\| \le \frac{\delta_1}{2}
\end{align*}
or
\begin{align*}
 \|\e^{\i \mathbf{H} T } - V_{\text{heat}}\|   \le \frac{\delta_1}{2 \|\bb{w}(0)\|.}
\end{align*}
According to Remark \ref{rem:errHom}, the error in one segment is
\[ \|\e^{\i \mathbf{H} \tau }-V_{\text{heat}}(\tau)\| \le C_{\text{heat}} \tau^2.\]
To suppress the error in the simulation over the total time $T$ to a small value $\delta_1/(2\|\bb{w}(0)\|)$, it suffices to divide the total time $T$ into $r$ intervals, where $r = T/\tau$, such that
\[\|(\e^{\i \mathbf{H} \tau })^r - (V_{\text{heat}}(\tau))^r\| \le r \|\e^{\i \mathbf{H} \tau }-V_{\text{heat}}(\tau)\|\le rC_{\text{heat}} \tau^2 \le \frac{\delta_1}{2\|\bb{w}(0)\|},\]
which implies $r \ge 2 C_{\text{heat}} T^2\|\bb{w}(0)\|/\delta_1$ queries to $O_{\mathbf{H}}$.

The estimate of $\varepsilon_1 = \|\tilde{\bb{w}}_b - \tilde{\bb{w}}_b^{aa}\|$ has been given in Lemma \ref{lem:inhomo}, with $\delta_1$ replaced by $\frac{\delta_1}{2}$. This suggests that we should choose
\[
K \ge \frac{ T^2}{\delta_1} \max\Big\{ 4 C_{\text{heat}} \|\tilde{\bb{b}}\|_{avg},\quad 2\sup_{t \in [0,T]} (\|\mathbf{H}\| \|\tilde{\bb{b}}(t)\| + \| \frac{\d}{\d t}\tilde{\bb{b}}(t)\|)  \Big\}.
\]
The proof is completed by combing Lemma \ref{lem:inhomo}.
\end{proof}

\subsection{Complexity analysis}

This section substitutes the quantities expressed in terms of $\tilde{\bb{w}}$ in the time complexity with those of $\bb{u}$. According to Lemma \ref{lem:map2w}, one has the following map
  \begin{align}
\ket{0} \otimes \ket{0^{n_s}} \otimes \ket{0^{n_{xp}}}
\quad \xrightarrow{} \quad
 \frac{1}{\eta_0 + \eta_1}\ket{0}\otimes \ket{0^{n_s}}
\otimes  \bb{w}^{aa}(T) + \ket{\bot}, \label{map2w1}
\end{align}
where $\bb{w}^{aa}(T)$ is the approximation of $\bb{w}(T)$.  For the heat equation, $H_1$ in \eqref{u2v} is negative semi-definite, which gives $\bb{u}(t)=\e^p \bb{v}(t,p)$ for all $p\ge 0$ and implies
\begin{equation}\label{wexpansion}
\bb{w}(t) = \sum_{i,k} \bb{v}_i (t,p_k) \ket{i,k} = \bb{u}(t) \otimes \sum_{p_k\ge 0} \e^{-p_k} \ket{k} + \sum_i \sum_{p_k< 0} \bb{v}_i (t,p_k)\ket{i,k}.
\end{equation}
For this reason, one can restore $\ket{\bb{u}^{aa}(T)}$, an approximation of $\ket{\bb{u}(T)}$, from \eqref{map2w1} by a projection onto $\ket{k}$ that satisfies $p_{k} >0$:
  \begin{align}
\frac{1}{\eta_0 + \eta_1}\ket{0}\otimes \ket{0^{n_s}}
\otimes  \bb{w}^{aa}(T) + \ket{\bot}
& \quad  \xlongequal{\mbox{normalisation}} \quad
\frac{\|\bb{w}^{aa}(T)\|}{\eta_0 + \eta_1}\ket{0}\otimes \ket{0^{n_s}}
\otimes  \ket{\bb{w}^{aa}(T)} + \ket{\bot}  \nonumber \\
& \quad  \xrightarrow{~~~\mbox{projection}~~~~} \quad
 \frac{\|\bb{w}^{aa}(T)\|}{\eta_0 + \eta_1}\ket{0}\otimes \ket{0^{n_s}}
\otimes  \ket{\bb{u}^{aa}(T)}  + \ket{\bot_1}, \label{statefinal}
\end{align}
with the probability given by
\[\text{P}_{\text{r}}^*
 = \frac{ \sum_{p_k>0} \|\bm{v}(T,p_k)\|^2}{\|\bm{w}(T)\|^2}
 = \frac{ \sum_{p_k>0} \|\bm{v}(T,p_k)\|^2}{\|\bm{w}(0)\|^2}
= \frac{C_{e0}^2}{C_e^2}\frac{\|\bb{u}(T)\|^2}{\| \bb{u}(0) \|^2} ,\]
where
\[C_{e0} = \Big(\sum_{p_k\ge 0}  \e^{-2 |p_k|} \Big)^{1/2}, \qquad C_e = \Big(\sum_{k=0}^{N_p-1} \e^{-2 |p_k|} \Big)^{1/2}.\]
The final state $\ket{\bb{u}^{aa}(T)}$ is obtained by measuring the state in \eqref{statefinal} and obtaining all 0 in the other qubits. The likelihood of acquiring this approximate state is
\begin{align*}
\text{P}_{\text{r}}^{meas}
& = \Big(\frac{\|\bb{w}^{aa}(T)\|}{\eta_0 + \eta_1}\Big)^2
= \Big(\frac{\|\bb{w}^{aa}(0)\|}{\eta_0 + \eta_1}\Big)^2 = \Big(\frac{\|\bb{w}(0)\|}{\|\bb{w}(0)\| + \|\tilde{\bb{b}}\|_{avg}}\Big)^2 \\
& = \frac{C_e^2 \|\bb{u}(0)\|^2}{( C_e \|\bb{u}(0)\| + C_e \|\bb{f}\|_{avg})^2 } = \frac{\|\bb{u}(0)\|^2}{( \|\bb{u}(0)\| + \|\bb{f}\|_{avg})^2 },
\end{align*}
where we have used the fact that
\[\|\tilde{\bb{b}}(t)\| =  \|\bb{b}(t)\| = \|\bb{f}(t) \otimes [\e^{-|p_0|}, \cdots, \e^{-|p_{N_p-1}|}]^T\| = C_e \|\bb{f}(t)\|, \]
\[\|\tilde{\bb{b}}\|_{avg} = \Delta s( \|\tilde{\bb{b}}(s_0)\| + \cdots + \|\tilde{\bb{b}}(s_{K-1})\| ) = C_e \|\bb{f}\|_{avg}.\]
Therefore, the overall probability is
\[\text{P}_{\text{r}} = \frac{C_{e0}^2}{C_e^2}\frac{\|\bb{u}(T)\|^2}{( \|\bb{u}(0)\| + \|\bb{f}\|_{avg})^2},\]
where
\[\frac{C_{e0}^2}{C_e^2} \approx \frac{\int_0^{\infty} \e^{-2p} \d p}{\int_{-\infty}^{\infty} \e^{-2 |p|} \d p} = \frac12. \]

\begin{theorem}\label{thm:overallu}
There exists a quantum algorithm that prepares a $\delta$-approximation of the state $\ket{\bb{u}(T)}$ with $\Omega(1)$ success
probability and a flag indicating success, using
\begin{enumerate}
  \item queries to the time evolution input oracle $O_{\mathbf{H}}$
  \[N_t^{LCU} = \Theta \Big(\frac{1}{Pr}  \frac{C_{\text{LCU}} T^2}{\delta} \Big),\]
  with
  \[\frac{1}{Pr}  =  \frac{C_e^3}{C_{e0}^2} \Big(\frac{\|\bb{u}(0)\| +\|\bb{f}\|_{avg}}{\|\bb{u}(T)\|}\Big)^3 \]
  and
   \[C_{\text{LCU}} =  C_{\text{heat}} +  \sup_{t \in [0,T]} (\frac{ N_p}{2R} \|A\| \|\bb{f}(t)\| +   \|\frac{\d }{\d t} \bb{f}(t)\|),\]
  where $C_{\text{heat}}$ is defined in Remark \ref{rem:errHom} and
  \[\| \bb{f}\|_{avg} = \Delta s( \|\bb{f}(s_0)\| + \cdots + \|\bb{f}(s_{K-1})\| );\]

  \item queries to the state preparation oracle $O_{\text{prep}}$, the source term input oracle $O_b$ and the coefficient oracle $O_{\text{coef}}$ for
  \[\mathcal{O}\Big( \frac{C_e^2}{C_{e0}^2} \Big(\frac{\|\bb{u}(0)\| +\|\bb{f}\|_{avg}}{\|\bb{u}(T)\|} \Big)^2 \Big)\]
  times.
\end{enumerate}
\end{theorem}
\begin{proof}
Using the inequality $\| \frac{x}{\|x\|} - \frac{y}{\|y\|} \| \le 2 \frac{\|x - y\|}{\|x\|}$ for two vectors \( x, y \), we can bound the error in the quantum state after a successful measurement as
\[
\| \ket{\bb{u}(T)} - \ket{\bb{u}^{aa}(T)} \| \le \frac{2 \| \bb{u}(T) - \bb{u}^{aa}(T) \|}{\| \bb{u}(T) \|} \le \delta,
\]
implying
\[
\| \bb{u}(T) - \bb{u}^{aa}(T) \| \le \frac{\delta \| \bb{u}(T) \|}{2}.
\]

Let \(\ket{k}\) denote the computational basis state corresponding to the successful measurement (i.e., \(p_k \geq 0\)). From Eq.~\eqref{wexpansion}, we have
\[
\bb{u}(T) = \e^{p_k} (I^{\otimes n_x} \otimes \bra{k}) \bb{w}(T), \qquad \bb{u}^{aa}(T) = \e^{p_k} (I^{\otimes n_x} \otimes \bra{k}) \bb{w}^{aa}(T).
\]
Given this, we can require
\[\| \bb{w}(T) - \bb{w}^{aa}(T) \| \le \frac{\delta \e^{-p_k} \| \bb{u}(T) \|}{2} .\]
In the real implementation, $p_k$ can be chosen as $\mathcal{O}(1)$, so we set $\delta_1 = \frac{\delta \| \bb{u}(T) \|}{2}$ for Lemma \ref{lem:map2w}.

It is simple to check that
 \[\|\bb{w}(0)\| = \|\bb{u}(0) \otimes [\e^{-|p_0|}, \cdots, \e^{-|p_{N_p-1}|}]^T\| = C_e \|\bb{u}(0)\|,\]
\[\|\frac{\d }{\d t}\tilde{\bb{b}}(t)\| =  \|\frac{\d }{\d t} \bb{b}(t)\| = \|\frac{\d }{\d t} \bb{f}(t) \otimes [\e^{-|p_0|}, \cdots, \e^{-|p_{N_p-1}|}]^T\| = C_e \|\frac{\d }{\d t} \bb{f}(t)\|, \]
and
\[\|\bb{H}\| =  \|A\| \cdot \|D_{\eta}\| = \frac{N_p}{2R} \|A\|.\]

Plugging above relations into Lemma \ref{lem:map2w} yields the desired result.
\end{proof}

\section{Autonomisation for inhomogeneous term}\label{sec:autonomization}

This section addresses the time-dependent inhomogeneous term of \eqref{generalSchrEnlarge} by using the autonomisation method proposed in \cite{CJL23TimeSchr}, where a non-autonomous system is transferred to an autonomous one in one higher dimension. One advantage of this approach is that it makes the quantum simulation of  non-autonomous system the same as the autonomous one, avoiding use of the  cumbersome Dyson's series.

\subsection{The autonomisation method}

\subsubsection{Autonomisation method for the enlarged system}

We first recall the time evolution operator. Let's consider the time-dependent linear system of ODEs
\begin{equation}\label{Hamiltoniansimulation}
\frac{\d }{\d t} \bb{y}(t) = -\i H(t) \bb{y}(t), \qquad \bb{u}(t_0) = \bb{y}_0,
\end{equation}
where the initial condition is provided at $t = t_0$. The solution to this equation defines a time evolution operator, $\mathcal{U}_{t,t_0}$, satisfying
\[\bb{y}(t) = \mathcal{U}_{t,t_0} \bb{y}_0.\]
Plugging it into \eqref{Hamiltoniansimulation} yields a differential equation for $\mathcal{U}_{t,t_0}$,
\[\frac{\d }{\d t} \mathcal{U}_{t,t_0} \bb{y}_0 = -\i H(t) \mathcal{U}_{t,t_0} \bb{y}_0.\]
Since $\bb{y}_0$ is arbitrary, it follows that $\mathcal{U}_{t,t_0}$ satisfies
\[\frac{\d }{\d t} \mathcal{U}_{t,t_0}  = -\i H(t) \mathcal{U}_{t,t_0} .\]
If the Hamiltonian $H(t)$ satisfies $[H(t_i), H(t_j)] = H(t_i)H(t_j) - H(t_j)H(t_i) = O$ for all choices of $t_i$ and $t_j$, then the operator can be rewritten as
\[\mathcal{U}_{t,t_0} = \exp\Big( -\i \int_{t_0}^t H(\tau) \d \tau \Big),\]
where the exponential of an operator is defined via its Taylor series. For the cases where $[H(t_i), H(t_j)] \ne O$, the evolution operator can be expanded as the Dyson series
\[\mathcal{U}_{t,t_0} = I + \sum_{n=1}^\infty (-\i)^n \int_{t_0}^t \d t_1  \int_{t_0}^{t_1} \d t_2 \cdots \int_{t_0}^{t_{n-1}} \d t_n H(t_1)H(t_2)\cdots H(t_n).\]
This may be compactly represented using the time-ordering operator $\mathcal{T}$ which sorts any sequence of $n$ operators
according to the times $t_j$ of their evaluation, that is,
\[\mathcal{T}[H(t_1)H(t_2)\cdots H(t_n)] = H(t_{i_1})H(t_{i_2})\cdots H(t_{i_n}), \qquad t_{i_1}> t_{i_2} > \cdots > t_{i_n}.\]
With this time-ordered product, the propagator is formally expressed as a time-ordered exponential \cite{Dyson1949Series,Dyson1949Smatrix} (see also \cite{low2018hamiltonian,Berry2019Dyson,BerryChilds2020TimeHamiltonian,CJL23TimeSchr}):
\[\mathcal{U}_{t,t_0} = \exp_{\mathcal{T}}\Big( -\i \int_{t_0}^t H(\tau) \d \tau \Big)
=: I + \sum_{n=1}^\infty \frac{(-\i)^n}{n!} \int_{t_0}^t \d t_1  \int_{t_0}^t \d t_2 \cdots \int_{t_0}^t \d t_n \mathcal{T}[H(t_1)H(t_2)\cdots H(t_n)].\]

\begin{theorem}\cite{CJL23TimeSchr,Peskin19943tt} \label{lem:autonomizationTheorem}
For the non-autonomous system in \eqref{Hamiltoniansimulation},  introduce the following initial-value problem of an autonomous PDE
\begin{align*}
& \frac{\partial \bb{z}}{\partial t} + \frac{\partial \bb{z}}{\partial s} = -\i H(s) \bb{z},\\
& \bb{z}(0,s) = G(s)\bb{y}_0, \qquad s\in \mathbb{R},
\end{align*}
where $G(s)$ is a single-variable function and $H(s) = 0$ if $s<0$.
The analytical solution to this problem is
\[\bb{z}(t,s) = G(s-t)\mathcal{U}_{s,s-t}\bb{y}_0,\]
where
\[
\mathcal{U}_{s,s-t} = \exp_{\mathcal{T}}\Big( -\i \int_{s-t}^s H(\tau) \d \tau \Big) = \exp_{\mathcal{T}}\Big( -\i \int_0^t H(s-t+\tau) \d \tau \Big).
\]
The solution to \eqref{Hamiltoniansimulation} can be expressed in terms of $\bb{y}$ as
\[\bb{z}(t,s = t) = G(0) \bb{y}(t), \qquad t\ge t_0. \]
\end{theorem}

Now we are ready to apply the above theorem to the enlarged problem \eqref{generalSchrEnlarge}, namely,
\[
\frac{\d}{\d t} \tilde{\bb{w}}_{\varepsilon}(t) = \i H_{\varepsilon}(t) \tilde{\bb{w}}_{\varepsilon}(t), \qquad
\tilde{\bb{w}}_{\varepsilon}(0) = \tilde{\bb{w}}_{\varepsilon,0},
\]
where $H_{\varepsilon}(t)$ can be decomposed as
\begin{align*}
H_{\varepsilon}(t)
& = - H_{1,\varepsilon}(t) \otimes D_\eta + H_{2,\varepsilon}(t) \otimes I  \\
& = \ket{0}\bra{0} \otimes H - \sigma_x \otimes \frac{\varepsilon B(t)}{2} \otimes D_\eta +
\sigma_y \otimes \frac{\varepsilon B(t)}{2} \otimes I \\
& = A_0 + B_{\varepsilon}(t),
\end{align*}
with
\[A_0 = \ket{0}\bra{0} \otimes H, \qquad H = - H_1 \otimes D_\eta + H_2 \otimes I, \]
\[B_{\varepsilon}(t) = - \sigma_x \otimes \frac{F_{\varepsilon}}{2} \otimes D_\eta +
\sigma_y \otimes \frac{F_{\varepsilon}}{2} \otimes I.\]
The corresponding autonomisation problem is formulated as
\begin{equation}\label{autoPDE}
\begin{cases}
 \dfrac{\partial \bb{z}}{\partial t} + \dfrac{\partial \bb{z}}{\partial s} = \i (A_0 + B_{\varepsilon}(s)) \bb{z}(t,s),\\
 \bb{z}(0,s) = G(s)\tilde{\bb{w}}_{\varepsilon,0}, \qquad s\in [-T,T].
\end{cases}
\end{equation}
For $s<0$, we define $B_{\varepsilon}(s) = O$, still ensuring $\bb{z}(t,s = t) = G(0) \tilde{\bb{w}}_{\varepsilon}(t)$ for $t\ge 0$. To simplify, we assume $T\ge 1$. Otherwise, one can rescale the function $G(s)$ defined below to ensure it has compact support within $[-T,T]$.
To apply the discrete Fourier transform to $s$, it is necessary to require that $\bb{z}(t,s)$ is periodic in the $s$ direction. To this end, we choose $G(s)$ as the mollifier defined by
\[G(s) = \begin{cases}
c \e^{1/(s^2-1)}, \qquad & |s|<1, \\
0, \qquad & \mbox{otherwise},
\end{cases}\]
where $c = e$ gives $G(0) = 1$.
This choice imposes the periodic boundary condition
\[\bb{z}(t,s=-T) = \bb{z}(t, s=T) = \bb{0}\]
in the $s$ direction.
To align with the formulation in Section \ref{sec:SchrLap}, we set $T = \pi R_s$ or $R_s = T/\pi$, thereby imposing the boundary condition
\[\bb{z}(t,s=-\pi R_s) = \bb{z}(t, s= \pi R_s) = \bb{0}.\]

In the sequel we assume that $A_0$ is a $(1+n_{xp})$-qubit matrix, where $n_{xp} = d n_x+n_p$.

\subsubsection{Discretisation of the autonomisation problem}

We choose a uniform mesh size $\Delta s = 2T/N_s = 2\pi R_s/N_s$ for the autonomisation variable with $N_s = 2^{n_s}$ being an even number,
with the grid points denoted by $-T = - \pi R_s =  s_0<s_1<\cdots<s_{N_s} = \pi R_s = T$. For ease of presentation, we arrange the collection of the function $\bb{z}$ at these grid points in a different order than that specified in Eq.~\eqref{orderw}. Specifically, we place the $s$-register in the first position and define
\[\bar{\bb{w}}(t) = \sum_{l=0}^{N_s-1}\sum_{i=0}^{N_x+N_p+1} \bb{z}_i(t,s_l) \ket{l,i}\]
where $\bb{z}_i$ is the $i$-th entry of $\bb{z}$.

By applying the discrete Fourier transformation in the $s$  direction, one arrives at
\begin{equation}\label{autodiscretization}
\frac{\d}{\d t} \bar{\bb{w}}(t) = \i \bar{\bb{H}} \bar{\bb{w}}(t), \qquad \bar{\bb{w}}(0) = \bar{\bb{w}}_0
= [G(s_0),\cdots, G(s_{N_s-1})]^T\otimes \tilde{\bb{w}}_{\varepsilon,0},
\end{equation}
where the Hamiltonian is
\begin{align*}
\bar{\bb{H}}
& = -P_s \otimes I^{\otimes (1+n_{xp})} + \sum_{k=0}^{M_s-1} \ket{l}\bra{l} \otimes (A_0 + B_{\varepsilon}(s_l))\\
& = -P_s \otimes I^{\otimes (1+n_{xp})} + I^{\otimes n_s} \otimes A_0  + \sum_{l=0}^{N_s-1}\ket{l}\bra{l} \otimes B_{\varepsilon}(s_l),
\end{align*}
with
\[P_s = F_s D_s F_s^{-1},  \qquad D_s = \text{diag}(\mu_0, \cdots, \mu_{N_s-1}),  \qquad \mu_l = \Big(l - \frac{N_s}{2}\Big) / R_s,\]
where $F_s$ is the matrix $\Phi$ given in \eqref{Pmu}.

\begin{remark}\label{zmeasure}
The unitary evolution gives
\[\|\bb{z}(T,s)\| = |G(s-T)| \cdot \|\tilde{\bb{w}}_{\varepsilon,0}\| ,\qquad s\in [-T,T].\]
For this reason, we can retrieve $\bb{z}(T,s=T) = \tilde{\bb{w}}_{\varepsilon}(T)$ with probability
\[\text{Pr}(s=T) = \dfrac{|G(0)|^2}{G^2(s_0)+\cdots + G^2(s_{N_s-1})} = : \dfrac{C_{G0}^2}{C_G^2}.\]
\end{remark}

Employing the first-order Lie-Trotter-Suzuki decomposition, we can approximate the time evolution operator for \eqref{autodiscretization} as
\begin{equation}\label{Uauto}
 \e^{\i \bar{\bb{H}} \tau} \approx U_{\text{auto}}^{(1)} (\tau) U_{\text{auto}}^{(2)}(\tau),
\end{equation}
where
\[U_{\text{auto}}^{(1)} (\tau) = \exp\Big( \i \tau (-P_s \otimes I^{\otimes (1+n_{xp})} + I^{\otimes n_s} \otimes A_0) \Big)\]
\[U_{\text{auto}}^{(2)}(\tau) = \exp\Big( \i \tau \sum_{l=0}^{N_s-1} \ket{l}\bra{l}  \otimes B_{\varepsilon}(s_l) \Big).\]
Here, the second operator incorporates the effects of time-dependent boundary conditions.

 As noted in Remark \ref{rem:logJ}, throughout the remainder of this article, we will assume \( \tau \le \tau_0 \), where \( \tau_0 \) is a small parameter such that \( J \tau \leq \tau_0 \). Here, \( J \) represents the total number of uses of a specific evolution operator. These parameters will be made clear later.

\subsection{Quantum circuit for the autonomised  problem}

\subsubsection{The time-independent term $U_{\text{auto}}^{(1)} (\tau)$} \label{subsec:Uauto1}

For $U_{\text{auto}}^{(1)} (\tau)$, a direct manipulation gives
\begin{align*}
U_{\text{auto}}^{(1)} (\tau)
& = (F_s \otimes I^{\otimes (1+n_{xp})}) \exp\Big( \i \tau (-  D_s \otimes I^{\otimes (1+n_{xp})}+ I^{\otimes n_s} \otimes  A_0 ) \Big) (F_s^\dag \otimes I^{\otimes (1+n_{xp})}) \\
& = (F_s \otimes I^{\otimes (1+n_{xp})}) \Big(\e^{-\i \tau D_s} \otimes \e^{\i \tau A_0}\Big) (F_s^\dag \otimes I^{\otimes (1+n_{xp})})\\
&  = (F_s \otimes I^{\otimes (1+n_{xp})})  \Big(  \e^{-\i \tau  D_s } \otimes  \e^{ \i \tau (\ket{0}\bra{0} \otimes H )}\Big) (F_s^\dag \otimes I^{\otimes (1+n_{xp})}) \\
& = (F_s \otimes I^{\otimes (1+n_{xp})})  \Big(I^{\otimes n_x} \otimes (\ket{0}\bra{0} \otimes \e^{ \i \tau  H }  + \ket{1}\bra{1}  ) \otimes  \e^{-\i \tau  D_s } \Big) (F_s^\dag \otimes I^{\otimes (1+n_{xp})}) \\
& =: (F_s \otimes I^{\otimes (1+n_{xp})})  \tilde{U}_{\text{auto}}^{(1)}  (F_s^\dag \otimes I^{\otimes (1+n_{xp})}),
\end{align*}
where $F_s$ is the quantum Fourier transform in the $s$ direction, and $\e^{ \i \tau H }$ is exactly the time evolution of the Schr\"odingerised Laplacian (see Section \ref{sec:SchrLap}). The diagonal unitary in $\tilde{U}_{\text{auto}}^{(1)} $ can be reformulated as
\begin{align*}
\e^{-\i \tau  D_s}
 = \sum\limits_{l=0}^{N_s-1}  \e^{-\i \tau (l - \frac{N_s}{2}) / R_s  } \ket{l}\bra{l}
 = ( \e^{-\i \tau   / R_s  } )^{-N_s/2}  \sum\limits_{l=0}^{N_s-1}  ( \e^{-\i \tau   / R_s  } )^l   \ket{l}\bra{l},
\end{align*}
where the multiplicative factors can be included by phase kickback described as follows.

\begin{figure}[!htb]
  \centering
  \includegraphics[width=\textwidth]{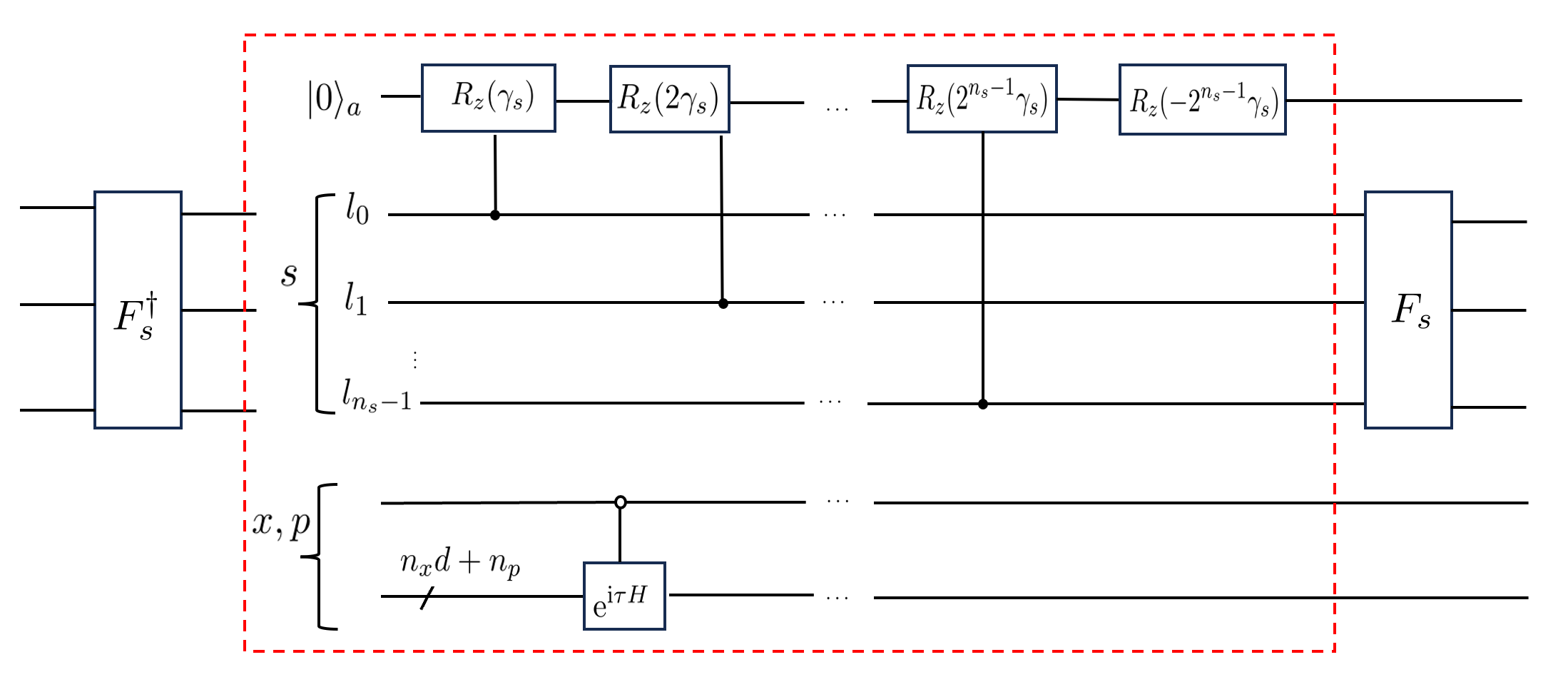}\\
  \caption{Quantum circuit for $U_{\text{auto}}^{(1)}(\tau)$. The dashed box corresponds to $\tilde{U}_{\text{auto}}^{(1)} (\tau)$. $F_s$ is the quantum Fourier transform in the $s$ direction. }\label{fig:URz}
\end{figure}

Let $R_z(\gamma_s)$ be the rotation gate with $\gamma_s = 2\tau/ R_s $. Introducing an ancilla qubit, we replace $\e^{-\i \tau  D_s} $ by the controlled-rotation
\[U_{cz}(\tau) = (I^{\otimes n_s} \otimes R_z^{-N_s/2}(\gamma_s) ) \sum\limits_{l=0}^{N_s-1}  \ket{l}\bra{l} \otimes R_z^l(\gamma_s)  ,\]
resulting in the quantum circuit in Fig. \ref{fig:URz}, where
     \begin{align*}
 \bullet \otimes\ket{0}
  \quad \xrightarrow{~~~U_{cz}(\tau)  ~~~} \quad
   \sum\limits_{l=0}^{N_s-1}   ( \e^{-\i \tau   / R_s  } )^{l-N_s/2} (\ket{l}\bra{l})  \bullet \otimes \ket{0},
   \end{align*}
where $\ket{0}$ corresponds to the ancilla qubit and we have used the simple relation $R_z(\theta) \ket{0} =  \e^{-\i \theta /2} \ket{0}$.
Thus, the desired state is obtained when the ancilla qubit is initialized in the 0 state. The select oracle can be realized by performing $c-R_z^{2^m}(\gamma_s)$ on the $s$-register controlled by the $m$-th qubit of the second register if it is 1.  From Lemma \ref{lem:selectH} we know that this select oracle can be constructed by $\log  N_s  = n_s$  queries to $R_z(\gamma_s )$.

Let the register for $s$ be $\ket{l}$, with the binary representation given by $l = (l_{n_s-1}, \cdots, l_0) = l_{n_s-1}2^{l_{n_s-1}}+ \cdots + l_02^0$. The quantum circuits representing $\tilde{U}_{\text{auto}}^{(1)} (\tau)$ and $U_{\text{auto}}^{(1)}(\tau)$ can be seen in Fig.~\ref{fig:URz}, where the dashed box corresponds to $\tilde{U}_{\text{auto}}^{(1)} (\tau)$.

\subsubsection{The time-dependent term $U_{\text{auto}}^{(2)} (\tau)$} \label{subsec:Uauto2}

For the second operator $U_{\text{auto}}^{(2)}(\tau) $ in \eqref{Uauto}, we obtain from the Taylor expansion of the matrix exponential that
\begin{equation}\label{Uauto2}
U_{\text{auto}}^{(2)}(\tau)  = \exp\Big( \i \tau \sum_{l=0}^{N_s-1} \ket{l}\bra{l} \otimes B_{\varepsilon}(s_l)\Big) = \sum_{l=0}^{N_s-1} \ket{l}\bra{l} \otimes \exp(\i \tau B_{\varepsilon}(s_l))  ,
\end{equation}
where $B_{\varepsilon}(s_l) = O$ for $l = 0,1,\cdots, N_s/2-1$. For $l\ge N_s/2$ one has
\begin{align*}
B_{\varepsilon}(s_l)
& = - \sigma_x \otimes \frac{F_{\varepsilon}(s_l)}{2} \otimes D_\eta +
\sigma_y \otimes \frac{F_{\varepsilon}(s_l)}{2} \otimes I^{\otimes n_p} \\
& = \sum_{k=0}^{N_p-1} (-\eta_k \sigma_x )\otimes \frac{F_{\varepsilon}(s_l)}{2} \otimes \ket{k}\bra{k}
+ \sigma_y\otimes \frac{F_{\varepsilon}(s_l)}{2} \otimes I^{\otimes n_p}.
\end{align*}
Employing the first-order Lie-Trotter-Suzuki decomposition, we can approximate $\exp(\i \tau B_{\varepsilon}(s))$ for $s\ge 0$ as
\begin{align}
\exp(\i \tau B_{\varepsilon}(s))
& \approx \exp\Big( \i \tau \sum_{k=0}^{N_p-1} (-\eta_k \sigma_x )\otimes \frac{F_{\varepsilon}(s)}{2} \otimes \ket{k}\bra{k} \Big) \exp\Big( \i \tau \sigma_y \otimes \frac{F_{\varepsilon}(s)}{2} \otimes I^{\otimes n_p} \Big) \nonumber \\
& =: \text{EXP}_x(\tau;s) \text{EXP}_y(\tau; s) . \label{TrotterUauto2}
\end{align}
For $s<0$, one can simply set $\text{EXP}_x(\tau;s) = \text{EXP}_y(\tau; s) = I$. A direct calculation gives
\begin{align} \label{appUauto2}
U_{\text{auto}}^{(2)}(\tau)
& \approx \Big(\sum_{l=0}^{N_s-1} \ket{l}\bra{l} \otimes \text{EXP}_x(\tau;s_l) \Big)
\Big(\sum_{l=0}^{N_s-1} \ket{l}\bra{l} \otimes \text{EXP}_y(\tau;s_l) \Big) \nonumber \\
& =: U_{\text{auto}}^{(2,x)}(\tau)  U_{\text{auto}}^{(2,y)}(\tau) .
\end{align}
Therefore, the final approximate evolution operator is
\[\e^{\i \bar{\bb{H}} \tau} \approx  U_{\text{auto}}^{(1)} (\tau) (U_{\text{auto}}^{(2,x)}(\tau)  U_{\text{auto}}^{(2,y)}(\tau)) =: U_{\text{auto}}(\tau) .\]

The Hamiltonian can be seen as evaluated at different discrete time steps, necessitating the introduction of a corresponding input model (the block-encoding used here) to coherently handle these terms. To this end, we briefly introduce the concept and the properties of the block-encoding that we will use in the sequel \cite{gilyen2019quantum}.

The core concept behind block-encoding is to represent the matrix $A$ as the upper-left block of a unitary matrix, i.e.,
\begin{equation*}
U_A \approx \begin{bmatrix} A & * & \cdots & * \\
* & * & \cdots & * \\
\vdots &  \vdots & \vdots & \vdots \\
*  &  * & \cdots & *
\end{bmatrix}  .
\end{equation*}

\begin{definition}\cite{Low2016qubitization,Low2017amplification,Chakraborty2018blockencode}
Given an $n$-qubit matrix $A$ satisfying $\|A\| \le \alpha$. If  $(n+n_a)$-qubit unitary matrix $U_A$  can be found such that
\[\|A - \alpha (\bra{0}_a \otimes I^{\otimes n}) U_A (\ket{0}_a \otimes I^{\otimes n})\| \le \varepsilon,\]
then $U_A$ is called the $(\alpha, n_a, \varepsilon)$-block-encoding of the matrix $A$.
\end{definition}

Given that each $B(s_l)$ is a diagonal matrix, we assume an exact block-encoding of
\[\bar{B} = \sum_{l=0}^{N_s-1} \ket{l}\bra{l} \otimes I \otimes  B(s_l) ,\]
where we add a single-qubit matrix $I$ for later uses. It is obvious that $\bar{B}$ remains a diagonal matrix, which can be realized from sparse Hamiltonian oracles (see \cite[Lemma 8]{Low2019Interaction} for example). The definition is given as follows.

\begin{definition}[Block encoding of time-dependent term]
Given a matrix $B_\varepsilon(s) = {F_{\varepsilon}(s)}/{2}: [-T,T] \to \mathbb{C}^{2^{n_x} \times 2^{n_x}}$ with $\|B_\varepsilon(t)\| \le \alpha$, assume that there exists an $(1+dn_x+n_s + n_a)$-qubit unitary oracle {\rm HAM} such that
\[{\rm HAM} = \begin{bmatrix} \bar{B}_\varepsilon/\alpha  & * \\ *  & *  \end{bmatrix}, \qquad
\bar{B}_\varepsilon = \sum_{l=0}^{N_s-1} \ket{l}\bra{l}\otimes I  \otimes B_\varepsilon(s_l) , \]
where $s_l = l \Delta s$ with $\Delta s = 2T/N_s$ and $n_a$ is the number of ancilla qubits for the block-encoding of $I\otimes \bar{B}_\varepsilon$. This means
\begin{equation}\label{projHAM}
(\bra{0}_a \otimes I^{\otimes (dn_x+n_s) }) {\rm HAM}  (\ket{0}_a \otimes I^{\otimes (dn_x+n_s)})
= \bar{B}_\varepsilon/\alpha.
\end{equation}
\end{definition}

In the implementation of \eqref{Uauto2}, we also require computing the function $\e^{\i t B}$ for a time-independent Hermitian matrix $B$. This can be achieved by first approximating $\cos(tB)$ and $\i \sin(tB)$ using even and odd polynomials, respectively, and then employing the standard quantum singular value transformation (QSVT) technique.

\begin{lemma}[Time-independent Hamiltonian simulation via QSVT, {\cite{Low2019Interaction,An2022timedependent}}] \label{lem:QSVT}
Let $\delta \in (0, 1)$, $t = \Omega(\delta)$ and let $U_H$ be an $(\alpha, n_a, 0)$-block-encoding of a time-independent Hamiltonian $B$. Then a unitary $V$ can be implemented such that $V$ is a $(1, n_a + 2, \delta)$-block-encoding of $\e^{\i t B}$, with $\mathcal{O}(\alpha t + \log (1/\delta))$ uses of $U_B$, its inverse or controlled version, $\mathcal{O}(n_a(\alpha t + \log (1/\delta)))$ two-qubit gates and $\mathcal{O}(1)$ additional ancilla qubits.
\end{lemma}

\begin{figure}[!htb]
  \centering
  \includegraphics[scale=0.2]{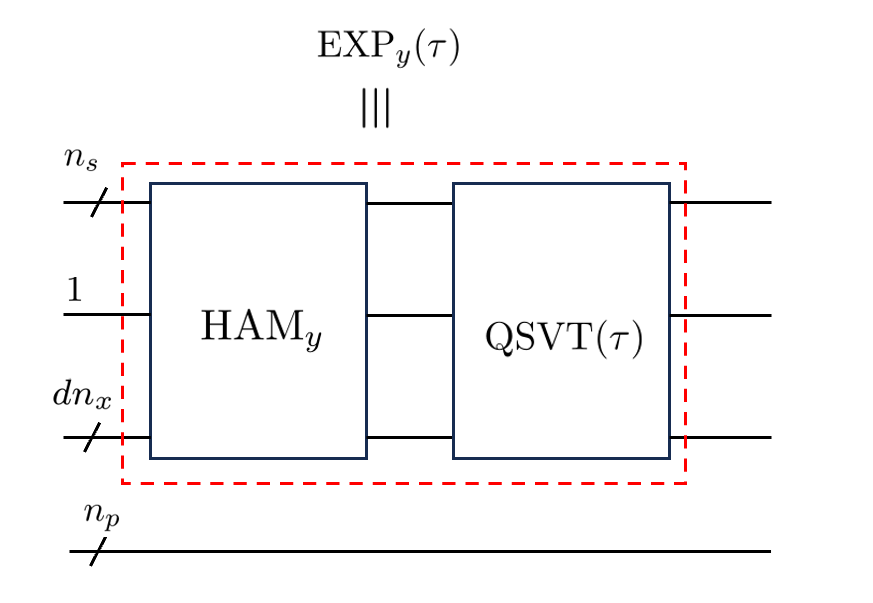}\\
  \caption{Quantum circuit for $U_{\text{auto}}^{(2,y)}(\tau)$. The dashed box corresponds to $\text{EXP}_y(\tau)$.}\label{fig:Uauto2y}
\end{figure}

For the second factor in \eqref{appUauto2}, it is straightforward to observe that the block encoding of
\[\sum_{l=0}^{N_s-1}  \ket{l}\bra{l} \otimes \sigma_y \otimes \frac{F_{\varepsilon}(s_l)}{2}\]
is $\text{HAM}_y:=(I^{\otimes n_s} \otimes \sigma_y \otimes I^{\otimes d n_x}) \text{HAM}$.
According to Lemma \ref{lem:QSVT}, utilizing the QSVT technique allows us to construct a $(1, n_a + 2, \delta_y)$-block-encoding of
\[\exp\Big( \i \tau \sum_{l=0}^{N_s-1} \ket{l}\bra{l} \otimes \sigma_y \otimes \frac{F_{\varepsilon}(s_l)}{2}\Big)=:\text{EXP}_y(\tau)\]
using $\mathcal{O}(\alpha \tau + \log(1/\delta_y) )$ applications of $\text{HAM}_y$, its inverse or controlled version,  $\mathcal{O}(n_a(\alpha \tau + \log(1/\delta_y) ) )$ two-qubit gates and $\mathcal{O}(1)$ additional ancilla qubits, where  $0< \tau \le \tau_0$ and $0<\delta_y\le\tau_0$ with $\tau_0$ to be determined. Then $U_{\text{auto}}^{(2,y)}(\tau)$ is obtained by adding the register in the $p$ direction, with the quantum circuit shown in Fig.~\ref{fig:Uauto2y}, where ancilla qubits used for block-encodings are not included.

\begin{figure}[!htb]
  \centering
  \includegraphics[width=\textwidth]{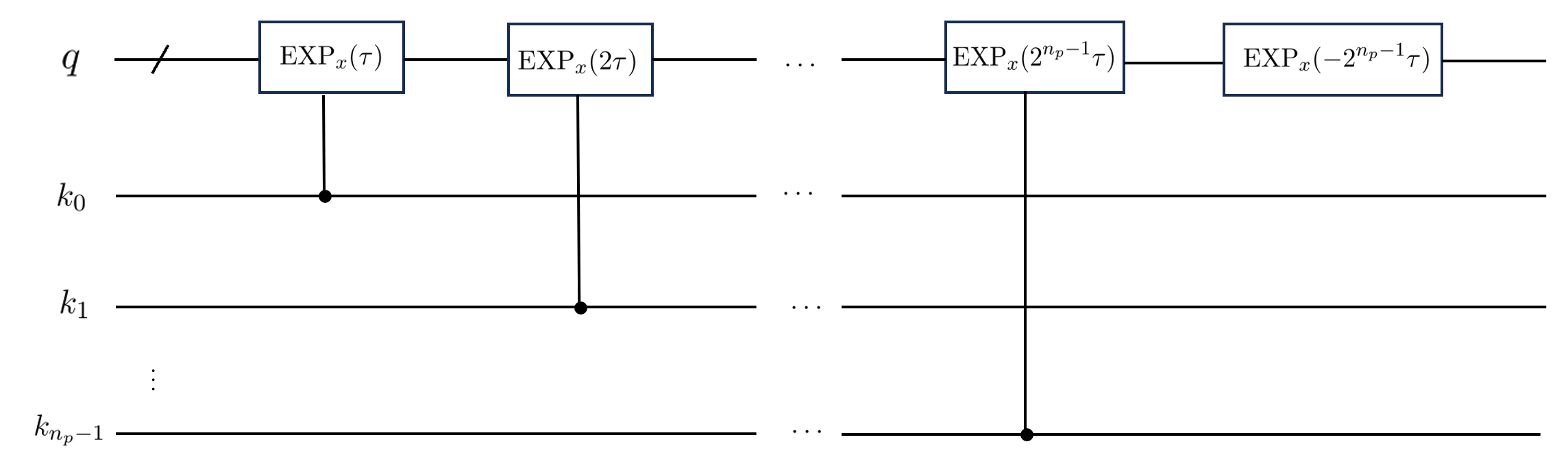}
  \includegraphics[scale=0.2]{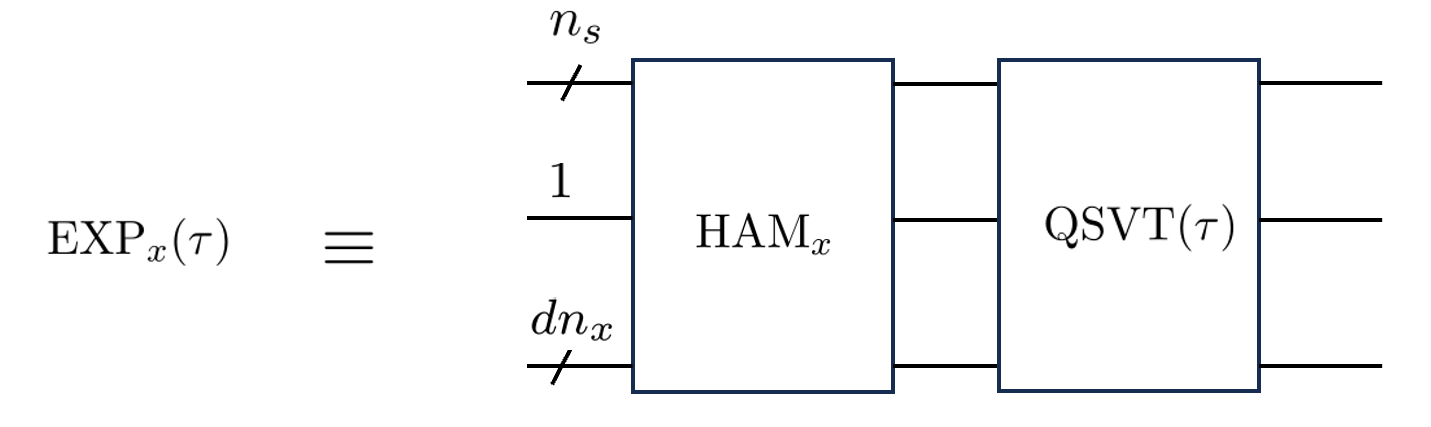}  \\
  \caption{Quantum circuit for $U_{\text{auto}}^{(2,x)}(\tau)$.}\label{fig:Uauto2x}
\end{figure}

For the first factor in \eqref{appUauto2}, we can reformulate $\text{EXP}_x(\tau;s)$ as
\begin{align*}
\text{EXP}_x(\tau;s)
& = \exp\Big( - \i\tau \sum_{k=0}^{N_p-1}  \eta_k\sigma_x \otimes \frac{F_{\varepsilon}(s)}{2} \otimes \ket{k}\bra{k} \Big) \\
& = \sum_{k=0}^{N_p-1}\exp\Big( - \i\tau/R \sigma_x \otimes  \frac{F_{\varepsilon}(s)}{2} \Big)^{k-N_p/2}\otimes  \ket{k}\bra{k} \\
& =: \sum_{k=0}^{N_p-1}V_x^{k-N_p/2}(\tau;s)\otimes  \ket{k}\bra{k}
  =  \Big(V_x^{-N_p/2}(\tau;s) \otimes I^{\otimes n_p} \Big) \sum_{k=0}^{N_p-1}V_x^k(\tau;s) \otimes  \ket{k}\bra{k},
\end{align*}
where
\[V_x(\tau;s) =\exp\Big( - \i\tau/R \sigma_x \otimes  \frac{F_{\varepsilon}(s)}{2} \Big), \qquad s\ge 0.\]
We also define $V_x(\tau;s) = I$ for $s<0$.  As before, the oracle $\text{HAM}_x:=(I^{\otimes n_s} \otimes \sigma_x \otimes I^{\otimes dn_x}) \text{HAM}$ block encodes
\[\sum_{l=0}^{N_s-1}\ket{l}\bra{l} \otimes \sigma_x \otimes \frac{F_{\varepsilon}(s_l)}{2}.\]
Then the QSVT technique gives a $(1, n_a + 2, \delta_x)$-block-encoding of
\begin{equation}\label{Vxsl}
\exp\Big( -\i \tau/R \sum_{l=0}^{N_s-1} \ket{l}\bra{l} \otimes \sigma_x \otimes \frac{F_{\varepsilon}(s_l)}{2}\Big)
= \sum_{l=0}^{N_s-1} \ket{l}\bra{l} \otimes V_x(\tau;s_l) =: \text{EXP}_x(\tau)
\end{equation}
using $\mathcal{O}(\alpha \tau/R + \log(1/\delta_x) )$ uses of $\text{HAM}_x$, its inverse or controlled version,  $\mathcal{O}(n_a(\alpha \tau/R + \log(1/\delta_x) ) )$ two-qubit gates and $\mathcal{O}(1)$ additional ancilla qubits.  We set $\tau_0 = \frac{N_p}{2R}\tau$ and $0<\delta_x \le \tau_0$, with $\delta_x$ and $\delta_y$ provided later. This allows constructing the following select oracle
\[\sum\limits_{k=0}^{N_p-1} \sum_{l=0}^{N_s-1} \ket{l}\bra{l} \otimes V_x^k(\tau;s_l) \otimes \ket{k}\bra{k} = \sum\limits_{k=0}^{N_p-1} \text{EXP}_x(k \tau) \otimes \ket{k}\bra{k} \]
using $\log J < n_p$ queries to \eqref{Vxsl}, where $J = N_p/2 = 2^{n_p-1}$.
The desired oracle $U_{\text{auto}}^{(2,x)}(\tau)$ is then obtained by applying $\text{EXP}_x(-2^{n_p-1} \tau)$ for $V_x^{-N_p/2}(\tau;s) \otimes I^{\otimes n_p}$, with the quantum circuit shown in Fig.~\ref{fig:Uauto2x}.

\begin{remark} \label{rem:countxy}
To sum up, we have
\begin{itemize}
  \item $U_{\text{auto}}^{(2,x)}(\tau)$ involves $\mathcal{O}(n_p( \alpha \tau /R + \log(1/\delta_x)) )$ uses to $\text{HAM}_x$, its inverse or controlled version and  $\mathcal{O}(n_an_p(\alpha \tau/R + \log(1/\delta_x) ) )$ two-qubit gates, with error $\mathcal{O}(n_p \delta_x)$, where $\delta_x \le \tau_0$.
  \item $U_{\text{auto}}^{(2,y)}(\tau)$ involves $\mathcal{O}(\alpha \tau + \log(1/\delta_y) )$ applications of $\text{HAM}_y$, its inverse or controlled version and $\mathcal{O}(n_a(\alpha \tau + \log(1/\delta_y) ) )$ two-qubit gates, with error $\mathcal{O}(\delta_y)$, where $\delta_y \le \tau_0$.
\end{itemize}
\end{remark}

\subsection{Complexity analysis}

For the discrete heat equation, the coefficient matrix $A$ associated with \eqref{ODElinear} is symmetric, leading to $ H_1 = A $ and $H_2 = O$ for the Schr\"odingerised system \eqref{generalSchr}. To bound the complexity from above, it is essential to carefully analyze the decomposition of the approximation error.

According to the theory of the Trotter splitting error with commutator scaling (referring to Eq.~(119) in \cite{Childs2021Commutator}), the splitting error for \eqref{Uauto} can be given by
\begin{align*}
& \|\e^{\i \bar{\bb{H}} \tau} - U_{\text{auto}}^{(1)} (\tau) U_{\text{auto}}^{(2)}(\tau)\| \\
& \le \frac{\tau^2}{2} \|[-P_s \otimes I^{\otimes (1+n_{xp})} + I^{\otimes n_s} \otimes \ket{0}\bra{0} \otimes (- H_1 \otimes D_\eta + H_2 \otimes I^{\otimes n_p})  ,  \\
& \hspace{4cm} \sum_{l=0}^{N_s-1} \ket{l}\bra{l}  \otimes (- \sigma_x \otimes \frac{F_{\varepsilon}(s_l)}{2} \otimes D_\eta +
\sigma_y \otimes \frac{F_{\varepsilon}(s_l)}{2} \otimes I^{\otimes n_p} ) ] \|\\
& \le \tau^2 \|-P_s \otimes I^{\otimes (1+n_{xp})} + I^{\otimes n_s} \otimes \ket{0}\bra{0} \otimes (- H_1 \otimes D_\eta + H_2 \otimes I^{\otimes n_p})\| \\
& \qquad \times \|\sum_{l=0}^{N_s-1} \ket{l}\bra{l}  \otimes (- \sigma_x \otimes \frac{F_{\varepsilon}(s_l)}{2} \otimes D_\eta +
\sigma_y \otimes \frac{\varepsilon B(s_l)}{2} \otimes I^{\otimes n_p} )\| \\
& \le \tau^2 ( \|P_s\| + \|A\| \cdot \|D_\eta\| ) \max_{l} \| (- \sigma_x \otimes \frac{F_{\varepsilon}(s_l)}{2} \otimes D_\eta +
\sigma_y \otimes \frac{F_{\varepsilon}(s_l)}{2} \otimes I^{\otimes n_p} )\| \\
& \le \tau^2 \Big( \frac{N_s}{2R_s} + \frac{N_p \|A\| }{2R}\Big) ( \frac{N_p }{2  R} + 1) \frac12 \max_{l} \|  F_{\varepsilon}(s_l)\| \\
& \le \Big( \frac{N_s \pi}{2 T} + \frac{N_p \|A\| }{2R}\Big)  ( \frac{N_p }{2  R} + 1) \frac12 g_f \tau^2,
\end{align*}
where
\[[A,B] = AB-BA \qquad \mbox{and} \qquad g_f = \max_{i,l} \frac{|f_i(s_l)|}{( (f_i^2)_{ave} + \varepsilon^2 )^{1/2}}.\]
In the implementation, one can choose $N_s$ such that $\frac{N_s \pi}{2 T} \le \frac{N_p \|A\| }{2R}$ (Note that $\|A\| = \mathcal{O}(d/\Delta x^2)$), which gives
\[\|\e^{\i \bar{\bb{H}} \tau} - U_{\text{auto}}^{(1)} (\tau) U_{\text{auto}}^{(2)}(\tau)\|
\le \frac{N_p \|A\| }{4 R}  ( \frac{N_p }{2  R} + 1)  g_f  \tau^2.\]
Note that we have disregarded the error arising from discretisation in the $s$ direction due to the spectral accuracy of the Fourier spectral method.

In the implementation of $U_{\text{auto}}^{(1)}(\tau)$ (see Fig.~\ref{fig:URz}), $\e^{\i \tau H}$ is approximated by the quantum circuit for the homogeneous problems, where $H$ is exactly the Schr\"odingerised Laplacian $\bb{H}_{\text{heat}}$. Let the approximated operator be denoted as  $U_{\text{auto}}^{(1,a)}(\tau)$. One easily finds that
\[\|U_{\text{auto}}^{(1)}(\tau) - U_{\text{auto}}^{(1,a)}(\tau)\| \le \|\e^{\i \tau \bb{H}_{\text{heat}}} - V_{\text{heat}}(\tau)\| \le C_{\text{heat}} \tau^2,\]
with $C_{\text{heat}}$ defined in Remark \ref{rem:errHom}.

For the operator $U_{\text{auto}}^{(2)}(\tau)$, we obtain from the Trotter splitting in \eqref{appUauto2} that
\begin{align*}
& \| U_{\text{auto}}^{(2)}(\tau)- U_{\text{auto}}^{(2,x)}(\tau)  U_{\text{auto}}^{(2,y)}(\tau)\| \\
& = \| \exp(\i \tau B_{\varepsilon}(s)) - \text{EXP}_x(\tau;s) \text{EXP}_y(\tau; s) \| \\
& \le \frac{\tau^2}{2}\| [ \sigma_x \otimes \frac{F_\varepsilon (s_l)}{2} \otimes D_\eta, ~~
\sigma_y \otimes \frac{F_\varepsilon (s_l)}{2} \otimes I^{\otimes n_p}  ] \| \\
& \le \frac{\tau^2}{2}\| [ \sigma_x, \sigma_y] \| \cdot \|\frac{F_\varepsilon (s_l)}{2}\|^2 \| D_\eta\| \\
& \le \frac{\tau^2}{2} \cdot 2 \cdot (\frac{1}{2} g_f  )^2 \cdot \frac{N_p}{2R} = \frac{N_p}{2R} (\frac{1}{2} g_f )^2 \tau^2 \le \frac{N_p}{8 R} g_f^2 \tau^2,
\end{align*}
where $U_{\text{auto}}^{(2,x)}(\tau)$ and $U_{\text{auto}}^{(2,y)}(\tau)$ are obtained from the block-encoding of the time-independent Hamiltonian simulation using the QSVT technique with
\begin{align*}
 \| U_{\text{auto}}^{(2,x)}(\tau) - U_{\text{auto}}^{(2,x,a)}(\tau)\| \lesssim n_p \delta_x, \qquad
 \| U_{\text{auto}}^{(2,y)}(\tau) - U_{\text{auto}}^{(2,y,a)}(\tau)\| \lesssim  \delta_y,
\end{align*}
where the superscript ``a'' represents an approximation. Let $U_{\text{auto}}^{(2,a)} (\tau) = U_{\text{auto}}^{(2,x,a)}(\tau)U_{\text{auto}}^{(2,y,a)}(\tau)$. We have
\begin{align*}
& \|  U_{\text{auto}}^{(2)} (\tau) - U_{\text{auto}}^{(2,a)} (\tau) \| \\
& \le \|  U_{\text{auto}}^{(2)} (\tau) - U_{\text{auto}}^{(2,x)}(\tau)  U_{\text{auto}}^{(2,y)}(\tau) \|
+ \|U_{\text{auto}}^{(2,x)}(\tau)  U_{\text{auto}}^{(2,y)}(\tau) - U_{\text{auto}}^{(2,x,a)}(\tau)U_{\text{auto}}^{(2,y,a)}(\tau)\| \\
& \le \frac{N_p}{8 R} g_f^2 \tau^2 + \| U_{\text{auto}}^{(2,x)}(\tau) - U_{\text{auto}}^{(2,x,a)}(\tau)\| + \| U_{\text{auto}}^{(2,y)}(\tau) - U_{\text{auto}}^{(2,y,a)}(\tau)\| \\
& \lesssim \frac{N_p}{8 R} g_f^2 \tau^2 + n_p \delta_x + \delta_y.
\end{align*}
When choosing
\begin{equation*}\label{deltaxy}
\delta_x = \delta_y = g_f^2 \tau^2 /R ,
\end{equation*}
we obtain
\[\|  U_{\text{auto}}^{(2)} (\tau) - U_{\text{auto}}^{(2,a)} (\tau) \| \lesssim  \frac{N_p}{ R} g_f^2 \tau^2.\]
Note that the condition $\delta_x, \delta_y \le \tau_0 = \frac{N_p}{2R}\tau$ will automatically be satisfied with the choice of $r = T/\tau$ as determined later in \eqref{Ntauto}.

Therefore, the overall error is
\begin{align}
& \|\e^{\i \bar{\bb{H}} \tau} - U_{\text{auto}}^{(1,a)} (\tau) U_{\text{auto}}^{(2,a)}(\tau)\| \nonumber\\
& \le  \|\e^{\i \bar{\bb{H}} \tau} - U_{\text{auto}}^{(1)} (\tau) U_{\text{auto}}^{(2)}(\tau)\|
  + \| U_{\text{auto}}^{(1)} (\tau) - U_{\text{auto}}^{(1,a)} (\tau) \|
  +  \|  U_{\text{auto}}^{(2)} (\tau) - U_{\text{auto}}^{(2,a)} (\tau) \| \nonumber\\
& \lesssim \frac{N_p \|A\| }{4 R}  ( \frac{N_p }{2  R} + 1) g_f \tau^2 + C_{\text{heat}} \tau^2 + \frac{N_p}{ R}  g_f^2 \tau^2 \nonumber\\
& \le \Big( \frac{N_p \|A\| }{ R}  ( \frac{N_p }{2  R} + 1) g_f + \frac{N_p}{ R}  g_f^2 + C_{\text{heat}} \Big) \tau^2 =: C_{\text{auto}} \tau^2 = \delta'. \label{errauto}
\end{align}
To suppress the error in the simulation over the total time $T$ to a small value $\delta$, it is sufficient to divide the total time $T$ into $r$ intervals, where $r = T/\tau$, such that
\begin{equation}\label{diffUauto}
\|(\e^{\i \bar{\bb{H}} \tau})^r - (U_{\text{auto}}^{(a)}(\tau)) ^r\| \le r \|\e^{\i \bar{\bb{H}} \tau} - U_{\text{auto}}^{(a)}(\tau)\|  \le r \delta' = \delta_1,
\end{equation}
where $U_{\text{auto}}^{(a)}(\tau) = U_{\text{auto}}^{(1,a)} (\tau) U_{\text{auto}}^{(2,a)}(\tau)$, yielding
\begin{equation}\label{Ntauto}
r \gtrsim \frac{C_{\text{auto}} T^2 }{\delta_1}.
\end{equation}

Since querying ${\rm HAM}_x$ and ${\rm HAM}_y$ can be equivalent to querying ${\rm HAM}$, we estimate the complexity in terms of the HAM-T oracle ${\rm HAM}$. It's worth noting that this oracle has a gate complexity comparable to that of the source term input oracle $O_b$ in the LCU method (see Eq.~\eqref{Obsourceinput}). Hereafter, we denote $\ket{\bar{\bb{w}}(T)}$ as the normalized solution of \eqref{autodiscretization}. Summarizing the preceding discussion, we are in a position to provide the complexity estimate described below.

\begin{lemma}\label{lem:barw}
Suppose $N_s$ is selected such that $ \frac{N_s \pi}{2 T} \le \frac{N_p \|A\| }{2R} $. Then there exists a quantum algorithm that prepares a $\delta_1$-approximation of the state $\ket{\bar{\bb{w}}(T)}$ with a failure probability of $ \mathcal{O}(\delta_1) $, at the cost of:
\begin{enumerate}
  \item $\mathcal{O}(r)$ queries to the time evolution input oracle $O_{\mathbf{H}}$,
  where
  \[r = \Theta\Big(\frac{C_{\text{auto}} T^2 }{\delta_1 } \Big), \]
  with
  \[C_{\text{auto}} = C_{\text{heat}} + \frac{N_p \|A\| }{ R}  ( \frac{N_p }{2  R} + 1) g_f + \frac{N_p}{ R}  g_f^2, \qquad g_f = \max_{i,l} \frac{|f_i(s_l)|}{( (f_i^2)_{ave} + \varepsilon^2 )^{1/2}},\]

  \item $\mathcal{O}(\alpha  (n_p/R+1) T +  N_t^{\text{auto}} n_p \log(1/\delta_1) )$ uses of the {\rm HAM} oracle, its inverse or controlled version,

  \item $\mathcal{O}(n_a(\alpha (n_p/R+1) T + N_t^{\text{auto}} n_p\log(1/\delta_1) ) )$ one- or two-qubit gates.
\end{enumerate}
\end{lemma}

\begin{proof}
Let $\ket{\bar{\bb{w}}^{(a)}(T)}$ be the approximate state. From \eqref{diffUauto} we have
\[\|\ket{\bar{\bb{w}}(T)} - \ket{\bar{\bb{w}}^{(a)}(T)}\|
\le \|(\e^{\i \bar{\bb{H}} \tau})^r - (U_{\text{auto}}^{(a)}(\tau)) ^r\| \| \ket{\bar{\bb{w}}(0)}\| \le \delta_1, \]
which means $\ket{\bar{\bb{w}}^{(a)}(T)} = (U_{\text{auto}}^{(a)}(\tau)) ^r \ket{\bar{\bb{w}}(0)}$ is a $\delta_1$-approximation of the state $\ket{\bar{\bb{w}}(T)}$.

The failure probability is described below.
From \eqref{errauto} we obtain $U^{(a)}_{\text{auto}}(\tau)$ is $\delta'$-close to the unitary operator $\e^{\i \bar{\bb{H}} \tau}$. The norm of $U^{(a)}_{\text{auto}}(\tau)$ is at least $(1-\delta')$. Therefore the failure
probability at each single step is bounded by $1 - (1-\delta')^2 \le 2 \delta'$. This leads the global
failure probability to be bounded by $1 - (1-2\delta')^{N_t^{\text{auto}}} = \mathcal{O}(N_t^{\text{auto}}\delta')$. Therefore, in order to bound the error and the failure probability by $\mathcal{O}(\delta_1)$, we can choose $N_t^{\text{auto}}\delta' \le \delta_1$. This precisely corresponds to the selection of $N_t^{\text{auto}}$ given in \eqref{Ntauto}.

The cost can be obtained from Remark \ref{rem:countxy}, where we have used the fact that $\delta_x = \delta_y \le \delta_1$.
\end{proof}

 \begin{theorem}\label{thm:overallauto}
Suppose \( N_s \) is selected such that $ \frac{N_s \pi}{2 T} \le \frac{N_p \|A\| }{2R} $. Then there exists a quantum algorithm that prepares a \(\delta\)-approximation of the state $\ket{\bb{u}(T)}$ with a failure probability of $ \mathcal{O}(\delta_1) $, where
\[\delta_1 = \frac{C_{G0} }{C_G} \frac{ \| \bb{u}(T) \|}{(\|\bb{u}(0)\|^2 + \|\bb{f}\|_{ave}^2+1)^{1/2}} \delta, \qquad
\dfrac{C_{G0}^2}{C_G^2} = \dfrac{|G(0)|^2}{G^2(s_0)+\cdots + G^2(s_{N_s-1})},\]
at the cost of:
\begin{enumerate}
  \item $\mathcal{O}( N_t^{\text{auto}} )$ queries to the time evolution input oracle $O_{\mathbf{H}}$,
  where
  \[N_t^{\text{auto}} = \Theta\Big(\frac{1}{\text{Pr}}\frac{C_{\text{auto}} T^2 }{\delta } \Big), \]
  with
  \[C_{\text{auto}} = \frac{N_p \|A\| }{ R}  ( \frac{N_p }{2  R} + 1) g_f + \frac{N_p}{ R}  g_f^2 + C_{\text{heat}}, \qquad g_f = \max_{i,l} \frac{|f_i(s_l)|}{( (f_i^2)_{ave} + \varepsilon^2 )^{1/2}}\]
  and
  \[\frac{1}{\text{Pr}} = \dfrac{\e^{2\lambda_+ T}C_G^3}{C_{G0}^2} \Big(\frac{\|\bb{u}(0) \| + \|\bb{f}\|_{ave}+1}{\|\bb{u}(T)\|} \Big)^3,\]
  where $\lambda_+$ is defined in Remark \ref{rem:plus}

  \item $\mathcal{O}(\alpha  (n_p/R+1) T +  N_t^{\text{auto}} n_p \log(1/\delta_1) )$ uses of the {\rm HAM} oracle, its inverse or controlled version,

  \item $\mathcal{O}(n_a(\alpha (n_p/R+1) T + N_t^{\text{auto}} n_p\log(1/\delta_1) ) )$ one- or two-qubit gates.
\end{enumerate}
 \end{theorem}
\begin{proof}
According to Remark \ref{zmeasure}, we can retrieve $\ket{\tilde{\bb{w}}_{\varepsilon}(T)}$ from $\ket{\bar{\bb{w}}(T)}$
with a probability given by
\[\text{Pr}_{\tilde{w}} = \dfrac{C_{G0}^2}{C_G^2} = \dfrac{|G(0)|^2}{G^2(s_0)+\cdots + G^2(s_{N_s-1})}.\]
Similar to \eqref{wexpansion}, one has
\begin{equation}\label{wexpansion1}
\bb{w}_{\varepsilon}(t) =  \bb{u}_\varepsilon(t) \otimes \sum_{p_k\ge \lambda_+ t} \e^{-p_k} \ket{k} + \sum_i \sum_{p_k< \lambda_+ t} \bb{v}_i (t,p_k)\ket{i,k}.
\end{equation}
To derive a state proportional to $\bb{u}_\varepsilon(T) = [\bb{u}(T); \bb{r}_\varepsilon(T)]$, we perform an inverse quantum Fourier transform, followed by a projection onto $\ket{p_k\ge \lambda_+ T}$ through measurement, where $\lambda_+$ is defined in Remark \ref{rem:plus}. The probability is
\[\text{Pr}_w = \frac{\sum_{p_k>0} \|\bm{v}(T,p_k)\|^2}{\|\bm{w}_\varepsilon(T)\|^2}{}
=  \frac{C_{e+}^2 \|[\bb{u}(t); \bb{r}_\varepsilon(t)]\|^2}{\|\bb{w}_\varepsilon(0) \|^2 }
= \frac{C_{e+}^2}{C_e^2} \frac{ \|\bb{u}(T) \|^2 + \|\bb{r}_\varepsilon(T)\|^2 }{\|\bb{u}(0) \|^2 + \|\bb{r}_\varepsilon(0)\|^2  },\]
where
\[C_{e+} = \Big(\sum_{p_k\ge \lambda_+ T}  \e^{-2 |p_k|} \Big)^{1/2}, \qquad
\frac{C_{e+}^2}{C_e^2} \approx \frac{\int_{\lambda_+ T}^{\infty} \e^{-2p} \d p}{\int_{-\infty} ^{\infty} \e^{-2|p|} \d p} = \frac{1}{2} \e^{-2 \lambda_+ T}. \]
 The final state $\ket{\bb{u}(T)}$ is obtained from $\bb{u}_\varepsilon(T)$ with probability
\[\text{Pr}_u = \frac{\|\bb{u}(T)\|^2}{\|\bb{u}(T) \|^2 + \|\bb{r}_\varepsilon(T)\|^2 }.\]
Therefore, the overall probability is
\[\text{Pr} = \frac{C_{e+}^2}{C_e^2} \frac{C_{G0}^2}{C_G^2} \frac{\|\bb{u}(T)\|^2}{\|\bb{u}(0) \|^2 + \|\bb{f}\|_{ave}^2+1}. \]

To obtain a $\delta$-approximation of the state $\ket{\bb{u}(T)}$, one can impose the condition
\[
\| \bb{u}(T) - \bb{u}^{(a)}(T) \| \le  \frac{\delta}{2} \| \bb{u}(T) \|.
\]
Considering that $\| \bb{u}(T) - \bb{u}^{(a)}(T) \| \le \| \bb{u}_\varepsilon(T) - \bb{u}_\varepsilon^{(a)}(T) \|$, we establish a stricter requirement
\[
\| \bb{u}_\varepsilon(T) - \bb{u}_\varepsilon^{(a)}(T) \| \le  \frac{\delta}{2} \| \bb{u}(T) \|.
\]
Let $\ket{k}$ denote the computational basis state corresponding to the successful measurement (i.e., \(p_k \geq \lambda_+ T\)). Similar to the proof of Theorem \ref{thm:overallu}, one can require
\[
\| \bb{w}_{\varepsilon}(T) - \bb{w}_{\varepsilon}^{(a)}(T) \| \le \frac{\delta \e^{-p_k} \| \bb{u}(T) \|}{2}
\]
and neglect the factor $\e^{-p_k}$.
Given that $\bb{z}(T,s=T) = \tilde{\bb{w}}_{\varepsilon}(T)$ and $\tilde{\bb{w}}_{\varepsilon}(T)$ represents the discrete Fourier transform of $\bb{w}_{\varepsilon}(T)$, it is sufficient to ensure that
\[
\| \bar{\bb{w}}(T) - \bar{\bb{w}}^{(a)}(T) \| \le \frac{\delta \| \bb{u}(T) \|}{2}.
\]
Noting that
\[
\| \ket{\bar{\bb{w}}(T)} - \ket{\bar{\bb{w}}^{(a)}(T)} \|
\le \frac{2 \| \bar{\bb{w}}(T) - \bar{\bb{w}}^{(a)}(T) \|}{\| \bar{\bb{w}}(T) \|}
\le \frac{\delta \| \bb{u}(T) \|}{\| \bar{\bb{w}}(T) \|}
= \frac{\delta \| \bb{u}(T) \|}{\| \bar{\bb{w}}(0) \|}
\]
and
\begin{align*}
\| \bar{\bb{w}}(0) \|
 = C_G \|\bb{w}_{\varepsilon,0}\| = C_G (\|\bb{u}(0)\|^2 + \|\bb{r}_\varepsilon(0)\|^2)^{1/2}
 = C_G (\|\bb{u}(0)\|^2 + \|\bb{f}\|_{ave}^2+1)^{1/2},
\end{align*}
we can require
\[\| \ket{\bar{\bb{w}}(T)} - \ket{\bar{\bb{w}}^{(a)}(T)} \| \le \frac{1}{C_G} \frac{ \| \bb{u}(T) \|}{(\|\bb{u}(0)\|^2 + \|\bb{f}\|_{ave}^2+1)^{1/2}} \delta \sim \frac{1}{C_G} \frac{ \| \bb{u}(T) \|}{\|\bb{u}(0)\| + \|\bb{f}\|_{ave}+1} \delta=:\delta_1.\]
The proof is finished by plugging it into Lemma \ref{lem:barw}.
\end{proof}

\subsection{Comparison with the LCU method}

To begin with the query complexity for the right-hand side term $\bb{f}(t)$, the LCU method utilizes the source term input oracle $O_b$ to encode $\bb{f}(t)$, while autonomisation employs the block-encoding oracle ${\rm HAM}$ associated with the coefficient matrix. Notably, the gate complexity of ${\rm HAM}$ is comparable to that of $O_b$ in the LCU method, prompting a comparison primarily based on their respective utilization frequencies.

In the LCU method, $\bb{f}(t)$ is integrated into a linear combination of unitary operators per term, with $O_b$ queried once per execution. Therefore, the algorithm accesses $O_b$ a single time per run, and the total query count depends on the measurement repetitions, independent of time marching steps.

Conversely, the autonomisation method encodes $\bb{f}(t)$ directly within the coefficient matrix, requiring a query to ${\rm HAM}$ at each time step. Consequently, ${\rm HAM}$ is queried repeatedly across each time step during execution, resulting in a total query count multiplied by the number of measurement repetitions.

Secondly, we examine the factors \( N_t^{\text{LCU}} \) and \( N_t^{\text{auto}} \) as presented in Theorems \ref{thm:overallu} and \ref{thm:overallauto} respectively. Both theorems share a similar factor \( \frac{\|\bb{u}(0)\| + \|\bb{f}\|_{\text{avg}}}{\|\bb{u}(T)\|} \). Notably, the term \( \frac{\e^{2\lambda_+ T}C_G^3}{C_{G0}^2} \) in Theorem \ref{thm:overallauto} might be larger than \( \frac{C_e^3}{C_{e0}^2} \) in Theorem \ref{thm:overallu}, especially when \( \lambda_+ \) is large. However, the term \( \|\bb{f}\| \) in \( N_t^{\text{LCU}} \) can be substantially greater than \( g_f = \max_{i,l} \frac{|f_i(s_l)|}{( (f_i^2)_{\text{ave}} + \varepsilon^2 )^{1/2}} \), even though the latter is scaled by an additional factor \( \frac{N_p}{R} \). For instance, in scenarios where the entries of \( \bb{f} \) change slowly (\( \lambda_+ = \mathcal{O}(1) \)) and \( T = \mathcal{O}(1) \), the autonomisation method may yield a smaller multiplicative factor since $\|\bb{f}\|$ may scale as $\mathcal{O}(N_x^{d-1})$.

\section{Conclusion} \label{sec:conclusion}

In this study, we have explored the explicit implementation of quantum circuits for solving the heat equation under physical boundary conditions. We presented two methods for handling the inhomogeneous terms arising from time-dependent physical boundary conditions.
The first method utilizes Duhamel's principle to express the solution in integral form, complemented by the application of the LCU technique for coherent state preparation. The second approach involves an augmentation technique to transform the inhomogeneous problem into a homogeneous one, followed by leveraging quantum simulation techniques described in \cite{CJL23TimeSchr} to handle the resulting non-autonomous system by transferring it to an autonomous system in a higher dimension. Detailed implementations of these methods were provided, accompanied by a comprehensive complexity analysis in terms of queries to the time evolution input oracle.

While our discussion has been centered on the heat equation, the methodologies presented are adaptable to a wide range of scenarios. These include the advection equation with inflow boundary conditions, the Fokker-Planck equation with no-flux boundaries, quantum dynamics incorporating artificial boundary conditions, and problems involving interface conditions. This versatility underscores the broader applicability and potential impact of quantum simulation in tackling complex differential equations across various domains.

\section*{Acknowledgements}
SJ and NL are supported by NSFC grant No. 12341104, the Shanghai Jiao Tong University 2030 Initiative and the Fundamental Research Funds for the Central Universities. SJ was also partially supported by  the Shanghai Municipal Science and
Technology Major Project (2021SHZDZX0102). NL also acknowledges funding from the Science and Technology Program of Shanghai, China (21JC1402900).
YY is supported by the National Science Foundation for Young Scientists of China (No. 12301561).


\newcommand{\etalchar}[1]{$^{#1}$}

\end{document}